\documentclass[twocolumn,showpacs,amsmath,amssymb,aps,physrev,superscriptaddress,floatfix]{revtex4-1}

\usepackage{lipsum}
\usepackage{multirow}

\usepackage[]{graphicx}
\usepackage[]{verbatim}
\usepackage[]{color}
\usepackage{overpic}
\usepackage{rotating}
\usepackage{mathtools}
\usepackage{appendix}
\usepackage{ulem}
\usepackage[margin=0.6in]{geometry}

\usepackage[]{hyperref}

 \hypersetup{
 colorlinks = true,
 linkcolor = blue,
 urlcolor = magenta,
 citecolor = magenta}

\usepackage{array}
\newcolumntype{L}[1]{>{\raggedright\let\newline\\\arraybackslash\hspace{0pt}}m{#1}}
\newcolumntype{C}[1]{>{\centering\let\newline\\\arraybackslash\hspace{0pt}}m{#1}}
\newcolumntype{R}[1]{>{\raggedleft\let\newline\\\arraybackslash\hspace{0pt}}m{#1}}
\newcolumntype{N}{@{}m{0pt}@{}}

\makeatletter
\newsavebox{\@brx}

\newcommand{\llangle}[1][]{\savebox{\@brx}{\(\m@th{#1\langle}\)}%
  \mathopen{\copy\@brx\mkern2mu\kern-0.8\wd\@brx\usebox{\@brx}}}
\newcommand{\rrangle}[1][]{\savebox{\@brx}{\(\m@th{#1\rangle}\)}%
  \mathclose{\copy\@brx\mkern2mu\kern-0.8\wd\@brx\usebox{\@brx}}}

  \newcommand{\lllangle}[1][]{\savebox{\@brx}{\(\m@th{#1\langle}\)}%
  \mathopen{\copy\@brx\copy\@brx\mkern4mu\kern-0.7\wd\@brx\usebox{\@brx}}}
\newcommand{\rrrangle}[1][]{\savebox{\@brx}{\(\m@th{#1\rangle}\)}%
  \mathclose{\copy\@brx\copy\@brx\mkern4mu\kern-0.7\wd\@brx\usebox{\@brx}}}

\graphicspath{{./}{./}}

\begin{document}
\title{Magnetic Anisotropy in Spin-3/2 with Heavy Ligand in Honeycomb Mott Insulators: Application to CrI$_3$}
\author{P. Peter Stavropoulos}
\affiliation{Department of Physics and Center for Quantum Materials, University of Toronto, 60 St.~George St., Toronto, Ontario, M5S 1A7, Canada}
\author{Xiaoyu Liu}
\affiliation{Department of Physics and Center for Quantum Materials, University of Toronto, 60 St.~George St., Toronto, Ontario, M5S 1A7, Canada}
\author{Hae-Young Kee}
\email{hykee@physics.utoronto.ca}
\affiliation{Department of Physics and Center for Quantum Materials, University of Toronto, 60 St.~George St., Toronto, Ontario, M5S 1A7, Canada}
\affiliation{Canadian Institute for Advanced Research, Toronto, Ontario, M5G 1Z8, Canada}

\begin{abstract}
Ferromagnetism in the two-dimensional CrI$_3$ has generated a lot of excitement, and 
it was recently proposed that the spin-orbit coupling (SOC) in Iodine may generate bond-dependent spin interactions leading to
magnetic anisotropy. Here we derive a microscopic spin model of S=3/2 on transition metals surrounded by heavy ligands
in honeycomb Mott insulators using a strong-coupling perturbation theory.
For ideal octahedra we find Heisenberg and Kitaev interactions,
which favor the magnetic moment along the cubic axis via quantum fluctuations. 
When a slight trigonal distortion of the octahedra is present together with the SOC, three additional interactions arise, comprised of 
the off-diagonal symmetric $\Gamma$ and $\Gamma^\prime$, and single-ion anisotropy. 
The resulting magnetic anisotropy pins the moment perpendicular to the honeycomb plane as observed in a single-layer of CrI$_3$, suggesting
the significance of SOC and trigonal distortion in understanding magnetism of two dimensional Mott insulators. 
Comparison to the spin-orbit coupled $J_{\rm eff}$= 1/2 and S=1 models is also presented.
\end{abstract}
\maketitle

\section{Introduction}

Transition metal trihalides (TMT) are layered materials composed of
transition metals (M) and halides (X) of the group 9 in a 1:3 ratio.
They have a honeycomb layered structure, and depending on the filling of the $d$-orbitals in the transition metals, some are semiconductors and some are metals.\cite{McGuire2017TM}
Among them, RuCl$_3$,
VI$_3$ 
and CrI$_3$ are  Mott insulators.
Magnetic orderings in these systems further establish the importance of electronic correlations
and call for a microscopic understanding of spin models.
For example, based on a strong-coupling perturbation theory of the generic spin model\cite{khaliullin2005orbital,jk2009prl,rau2014prl},
 it was shown that $\alpha$-RuCl$_3$  described by the effective spin $J_{\rm eff}=1/2$ has dominant bond-dependent Kitaev and off-diagonal symmetric $\Gamma$ interactions.\cite{plumb2014prb,HSKim2015prb}
RuCl$_3$ has become an emergent candidate of spin-1/2 Kitaev spin liquid\cite{kitaev2006}.
Intense research activities on various properties of RuCl$_3$ have been carried out\cite{sandilands2015continuum,HSKim2015prb,sears2015prb,johnson2015monoclinic,banerjee2016proximate,cao2016structure,HSKim2016structure,janssen2017model} and recently a magnetic-field induced spin liquid was suggested.\cite{yadav2016field,baek2017evidence,wolter2017induced,zheng2017gapless,jansa2018types,kasahara2018thermal,yamashita2020sample}
 
In parallel theoretical interest in the ground state of higher-spin Kitaev models was initiated by classical model studies.\cite{Baskaran2008SpinS,Oitmaa2018SpinS} 
The classical Kitaev model has a macroscopic degeneracy named a classical spin liquid\cite{Baskaran2008SpinS},
but the higher-spin quantum Kitaev model is not exactly solvable, and the ground state is currently unknown. 
Various numerical studies such as exact diagonalization on S=1 suggested that the ground
state is possibly a spin liquid with gapless excitations.\cite{Nasu2018S=1}
These studies were mainly of theoretical interest, until a microscopic derivation of the S=1 Kitaev-Heisenberg model in multi-orbital systems was
found.\cite{Peter2019HigherK} Heavy ligand spin-orbit coupling (SOC) and strong Hund\rq{}s coupling in $e_{\rm g}$ orbitals
is a way to generate S=1 bond-dependent Kitaev interaction.
The magnetic field effects on the S=1 Kitaev model have also been investigated.\cite{Hickey2020S=1,Zhu2020S=1,Khait2020S=1}

The bond-dependent interactions have recently been adopted into TMT systems, 
because the nearest neighbor (n.n.) Heisenberg $J$, Kitaev $K$ and $\Gamma$ interactions are allowed based on the symmetry of the lattice.\cite{rau2014prl,YamajiPRL2014,KatukuriNJP2014}
In particular, ferromagnetism in a single-layer CrI$_3$ has generated excitement in recent years. 
\cite{HangJMCC2015,
TorelliIOP2018,
AlexanderPRC2018,ZhengNANO2018,
WebsterPRB2018,
JiwuAIP2018,
WuNC2019,
olsenMRS2019,
IgorPRB2019,
KumarIOP2019,
ChangsongPRL2020,
OlegJPC2020,
PizzocheroIOP2020,
NunezPRB2020,Soriano2020CrX3}
CrI$_3$ is a ferromagnetic (FM) insulator with $T_c\sim$ 61K for bulk samples.\cite{Handy1952JoACS,Dillon1965CrI3,McGuire2015CoM} 
Single-layer CrI$_3$ was successfully synthesized, which showed an FM ordering with $T_c\sim$ 45K. \cite{Huang2017MonolayerCrI3} 
The two-dimensional FM Heisenberg model is insufficient to explain finite $T_c$, i.e., the Mermin-Wagner theorem\cite{MerminWagnerPRL1966},
and several theoretical models were proposed to explain the magnetic anisotropy.
They include the XXZ model\cite{Lado2017XXZ,Kim2019XXZ}, single-ion anisotropy and Kitaev\cite{Xu2018Kitaev},
and large Kitaev and small symmetric off-diagonal $\Gamma$ interactions\cite{Lee2020PRL}.

 While the Heisenberg, Kitaev, and $\Gamma$ interactions are allowed by the symmetry, and found to be significant in 
the earlier derivations for lower-spins\cite{jk2009prl,rau2014prl,Peter2019HigherK}, their strengths may not be significant in S=3/2 systems.
Thus, a microscopic derivation of S=3/2 model is necessary to find the sources of the magnetic anisotropy.
 Here we derive a n.n. spin model for
S=3/2 with three electrons in $t_{\rm 2g}$ orbitals of transition metal sites and strong SOC in $p$-orbitals of ligands.
We take into account 
strong electron-electron interactions in multi-orbital systems including Hund\rq{}s coupling, and effects of trigonal distortions present in $R{\bar 3}$ 
rhombohedral lattice.
Contributions from $e_{\rm g}$ orbitals are important as shown below.
The minimal n.n. model includes $J$, $K$, $\Gamma$, another symmetric off-diagonal $\Gamma^\prime$\cite{rau2014trigonal}, and single-ion anisotropy $A_c$ along the $\hat{c}$-axis,
denoted as the $J-K-\Gamma-\Gamma^\prime-A_c$ model.

The rest of the paper is organized as follows. In Sec.~\ref{sec:kanamoriTightBinding}, the on-site Kanamori interaction and tight binding Hamiltonian are presented. 
In Sec.~\ref{sec:ideal}, we derive the n.n. spin model consisting of Kitaev and Heisenberg interactions for ideal octahedra environment using standard perturbation theory. 
In Sec.~\ref{sec:distortions}, we study the spin model  with trigonal distortion present in $R{\bar 3}$ rhombohedral lattice.
This includes the distortion-induced hopping matrix elements and three additional spin interactions generated via combined effects of SOC and distortion.
In Sec.~\ref{sec:dft}, we apply the theory to CrI$_3$ and present the exchange interaction strengths using tight binding parameter sets obtained by density functional theory.
The effects of the resulting magnetic anisotropy on the moment direction are found in Sec.~\ref{sec:maganiso}.
In Sec.~\ref{sec:spingap} we discuss the origin of the spin gap, finite $T_c$, and spin wave spectrum within
the $J-K-\Gamma-\Gamma^\prime-A_c$ model. The effects of the
the second n.n. Dzyaloshinskii-Moriya (DM) interaction is also discussed.
Finally in Sec.~\ref{sec:summary}, we summarize our results and compare with $J_{\rm eff}$=1/2 and S=1 spin models.
The detailed calculations
are presented in the Appendix.

\section{Kanamori interaction and tight binding Hamiltonian}\label{sec:kanamoriTightBinding}

The honeycomb network is made of metal (M) $d$-orbital sites with half filled $t_{\rm 2g}$ orbitals and octahedra cages of non-magnetic ligand (X) sites with fully occupied $p$-orbitals. 
The full Hamiltonian is composed of the on-site Kanamori interaction and the tight binding Hamiltonian between two sites.

\begin{figure}
  \centering
  \includegraphics[width=0.9\linewidth]{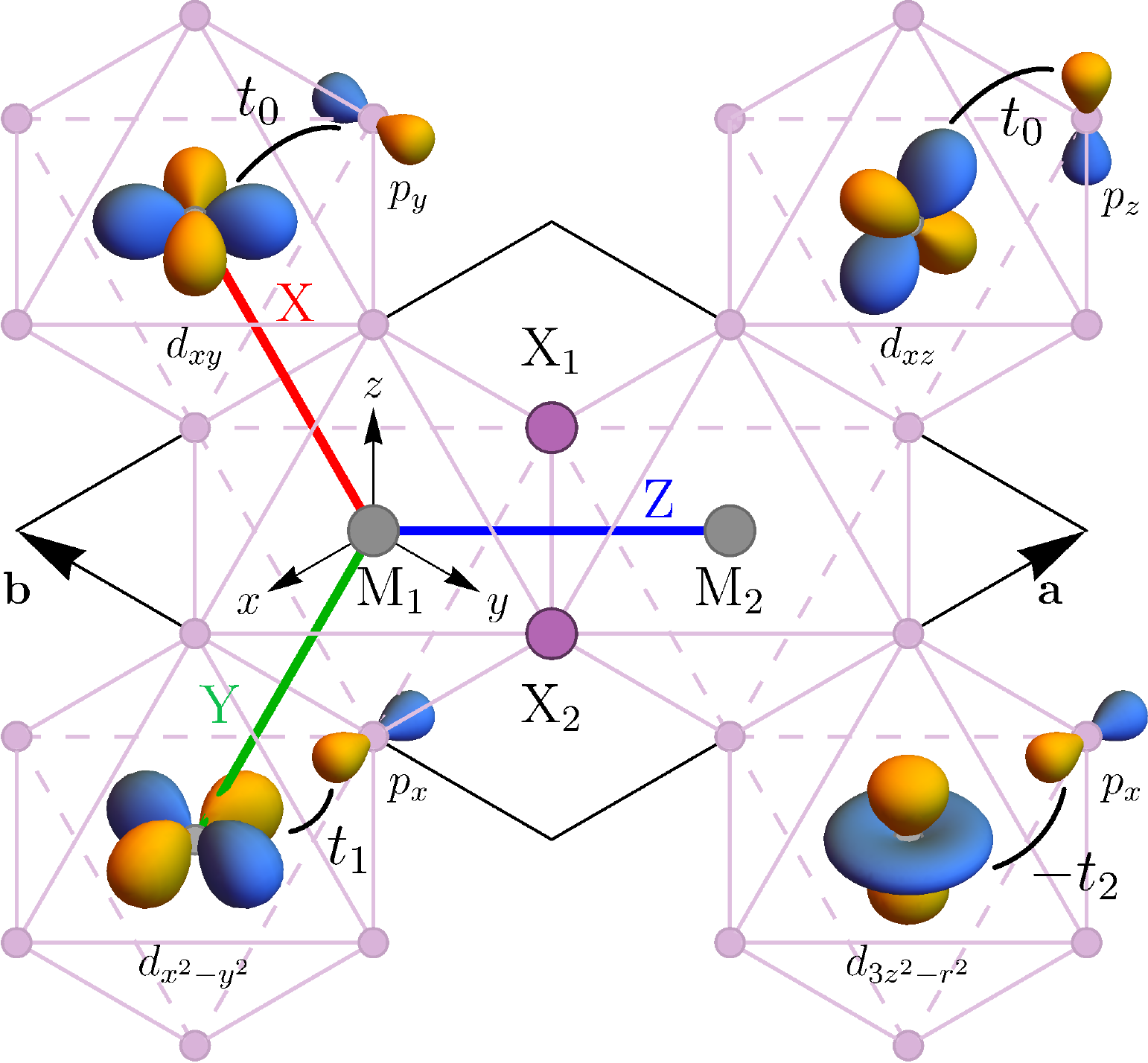}\\
  \caption{Edge-shared octahedra honeycomb structure unit cell $\mathbf{a,\ b}$, in global coordinates $xyz$. Transition metal sites M in gray and non-magnetic ligands X in purple. The n.n. bonds X, Y, Z are related by $C_3$ symmetry. The sites M$_1$, M$_2$, X$_1$, and X$_2$ are involved in the second order strong-coupling expansion on the Z bond. Indirect hopping integrals $t_0$, $t_1$ and $t_2$ are shown.}
  \label{fig:conventions}
\end{figure}

The on-site Hamiltonian of the  M sites is described by the Kanamori interaction~\cite{Kanamori1963PoTP} as well as crystal field spitting:
\begin{eqnarray}
H_{ee}& =  &  U \sum\limits_{\alpha} n_{\alpha\uparrow} n_{\alpha\downarrow} + \dfrac{U'}{2} \sum\limits_{\substack{\alpha\neq\beta,\\\sigma,\sigma'}} n_{\alpha\sigma} n_{\beta\sigma'} \nonumber \\
& & - \dfrac{J_{H}}{2}\sum\limits_{\substack{\alpha\neq\beta,\\\sigma,\sigma'}} c^{\dagger}_{\alpha\sigma}  c^{\dagger}_{\beta\sigma'} c_{\beta\sigma} c_{\alpha\sigma'}
+ J_{H}\sum\limits_{\alpha\neq\beta} c^{\dagger}_{\alpha\uparrow} c^{\dagger}_{\alpha\downarrow} c_{\beta\downarrow} c_{\beta\uparrow}\nonumber \\
& & + \Delta_c\sum\limits_{\substack{\alpha\in e_{\rm g},\\\sigma}} c^{\dagger}_{\alpha\sigma} c_{\alpha\sigma},
\label{eq:kanamori}
\end{eqnarray}
where the density operator $n_{\alpha\sigma}$ is given by $ c^{\dagger}_{\alpha \sigma} c_{\alpha \sigma}$, and
$c^{\dagger}_{\alpha\sigma}$ is the creation operator with $\alpha$ orbital and spin $\sigma$. 
$U$ and $U'$ are the intra-orbital and inter-orbital Hubbard interaction respectively, and $J_{H}$ is the Hund's coupling for the spin-exchange and pair-hopping terms. $\Delta_c$ is a crystal field splitting on the M sites, originated from the surrounding octahedra, leading to
the splitting of the $d$-orbitals into $t_{\rm 2g}$ and $e_{\rm g}$ orbitals. 
In a $d^3$ system one has half-filled $t_{\rm 2g}$ orbitals, where the Hund's coupling $J_H$ selects for the S=3/2 configuration as the ground state, and the angular momentum
is quenched.
A table of the excited state energy spectrum is show in Appendix~\ref{apx:spectrume}. 
The energies of the exited states are larger than the hopping integrals, which allows us to treat 
the tight binding hopping integrals as a perturbation.

In the edge shared octahedra structure, each bond between n.n. M sites involves two adjacent ligands, as shown in Fig.~\ref{fig:conventions}. A tight binding Hamiltonian between two transition metal sites M$_1$ and M$_2$ including the two adjacent ligands X$_1$ and X$_2$ 
is given below
\begin{equation}
H_{TB}=
\left(
\arraycolsep=2pt\def\arraystretch{1.5}
\begin{array}{llll}
 \mathrm{0}_{5\times5} & \mathbf{T_{M_1M_2}} & \mathbf{T_{M_1X_1}} & \mathbf{T_{M_1X_2}}\\
 \mathbf{T_{M_1M_2}}^{\dagger} & \mathrm{0}_{5\times5} &   \mathbf{T_{M_2 X_1}} & \mathbf{T_{M_2 X_2}}\\
\mathbf{ T_{M_1X_1}}^{\dagger} & \mathbf{T_{M_2X_2}}^{\dagger} & \mathrm{0}_{3\times3} & \mathrm{0}_{3\times3} \\
 \mathbf{T_{M_1X_2}}^{\dagger} & \mathbf{T_{M_2X_2}}^{\dagger} & \mathrm{0}_{3\times3} & \mathrm{0}_{3\times3}
\end{array}
\right),
\end{equation}
where $0_{n\times n}$ refers the $n\times n$ null matrix.
The basis is chosen as $( C_{\mathrm{M}_1,d}^{\dagger},C_{\mathrm{M}_2,d}^{\dagger},C_{\mathrm{X}_1,p}^{\dagger},C_{\mathrm{X}_2,p}^{\dagger})$, where $ C_{\mathrm{M}_i,d}^{\dagger} =
\left( c^{\dagger}_{i,x^2-y^2}, c^{\dagger}_{i,3z^2-r^2},  c^{\dagger}_{i,yz},  c^{\dagger}_{i,xz},  c^{\dagger}_{i,xy}, \right)$ are five $d$-orbitals at site $\mathrm{M}_i$, and $C_{\mathrm{X}_m,p}^{\dagger} = \left(   c^{\dagger}_{m,p_x},  c^{\dagger}_{m,p_y},  c^{\dagger}_{m,p_z} \right)$  are three $p$-orbitals at ligand site $\mathrm{X}_m$.
Each block of indirect hopping between $\mathrm{M}_i$ and $\mathrm{X}_m$ is denoted by $\mathbf{T_{M_i X_m}}$ and
the direct hopping between $\mathrm{M}$ sites by $\mathbf{T_{M_1 M_2}}$. The details of each block matrix will be presented latter.

To account for the indirect $d$ to $p$ hoppings we integrate out the $p$-orbitals through a perturbative procedure truncated at second order, leading to an effective $d$ to $d$ hopping model:
\begin{equation}\label{eq:integrateout}
\mathbf{T^{\mathrm{eff}}_{M_iM_j}} = \sum_{(a,m)} \dfrac{\mathbf{T_{M_iX_m}}| a \rangle \langle a |\mathbf{T_{X_m M_j}} }{\Delta E_{a}},
\end{equation}
where $(a,m)$ represent a sum over all single hole states $a$ of all sites X$_m$. The hole states are SOC states, thus creating two energy costs $\Delta E_{a}= \Delta-\lambda_p/2 \text{ or } \Delta+\lambda_p $, where $\Delta=\epsilon_d-\epsilon_p$ is the atomic energy difference between M and X sites, and $\lambda_p$ is the SOC in $p$-orbitals. The SOC will introduce explicit spin dependence in the effective $d$ to $d$ hopping. The total effective hopping between the two M sites now reads
\begin{equation}\label{eq:effective_total_hop}
H^{\rm eff}_{TB}=
\left(
\arraycolsep=2pt\def\arraystretch{1.5}
\begin{array}{llll}
 \mathrm{0}_{10\mathrm{x}10} & \mathbf{T_{M_1M_2}} + \mathbf{T^{\mathrm{eff}}_{M_1M_2}} \\
 \mathbf{T_{M_1M_2}}^{\dagger} + \mathbf{T^{\mathrm{eff}}_{M_1M_2}}^{\dagger} & \mathrm{0}_{10\mathrm{x}10} 
\end{array}
\right),
\end{equation}
where we still retain the bare direct hoppings $\mathbf{T_{M_1M_2}}$. Below we focus on the ideal honeycomb structure and first examine the effects of $\mathbf{T^{\mathrm{eff}}_{M_1M_2}}$ before adding the direct hoppings $\mathbf{T_{M_1M_2}}$ and summarizing the resulting spin model.

\section{Ideal honeycomb structure}\label{sec:ideal}
To understand the microscopic origin of the spin model, we start with the ideal honeycomb network surrounded by perfect edge shared octahedra.
It was shown that the
symmetry of the edge-shared octahedra Z bond
allows Heisenberg $J$, Kitaev $K$, and symmetric off-diagonal $\Gamma$ interactions~\cite{rau2014prl,YamajiPRL2014,KatukuriNJP2014}.
However, since their strengths depend on various exchange processes, we perform
the strong coupling perturbation theory to determine the exchange terms.


Truncating at second order in perturbation theory, we arrive at the following Heisenberg-Kitaev $(J-K)$ spin model for the ideal honeycomb octahedra. 
\begin{equation}\label{eq:spin_model_ideal}
\begin{array}{c}
H =\displaystyle \sum\limits_{\langle ij \rangle \in \gamma} J_0 {\bf S}_i \cdot {\bf S}_j + K_0 S_i^\gamma S_j^\gamma, \\[0.4cm]
\end{array}
\end{equation}
where $\gamma = x,y,z$ bond, and $J_0$ and $K_0$ refer to Heisenberg and Kitaev interactions for the ideal octahedra. 


 Below we present the details of the derivation of Heisenberg and Kitaev interactions. 
An explanation of the absence of the $\Gamma$ interaction within the second order perturbation theory is also discussed. 
The exchange processes include the contributions from both indirect and direct hoppings. 
We focus on the Z bond of the honeycomb Fig.~\ref{fig:conventions}, as the other two bonds are related by $C_3$ symmetry.

\subsection{Superexchange path: indirect hopping}\label{sec:superexchangeIndirect}

We first consider indirect hoppings between the M and X sites, which are the largest hopping integrals. They enter through the effective hoppings of Eq.~(\ref{eq:effective_total_hop}). The non-zero indirect hopping between $\mathrm{M}_1$ and $\mathrm{X}_1$ sites $t_0,\ t_1,\ t_2$ are shown in Fig.~\ref{fig:conventions}. These hoppings are 
incorporated in $\mathbf{T^{eff}_{M_1M_2}}$, which can be simplified using
the Slater-Koster decomposition and symmetry related $M-X$ bonds as shown in the Appendix. 
There are two contributions to both Heisenberg and Kitaev interactions, i.e., each interaction is composed of two exchange terms; one is from $t_{2g}-t_{2g}$ hoppings and the other is from $e_{\rm g}-t_{\rm 2g}$ hoppings 
\begin{equation}
J_0 = J_0^{\rm t_{2g}} + J_0^{\rm e_{g}},\ K_0 = K_0^{\rm t_{2g}} + K_0^{\rm e_{g}}, 
\end{equation} 
where the superscript ${\rm t_{2g}}$ and ${\rm e_g}$ refer to its corresponding hopping processes. 
Below we present each exchange path leading to $J_0^{\rm t_{2g}}$,  $K_0^{\rm t_{2g}}$, $J_0^{\rm e_{g}}$, and $K_0^{\rm e_{g}}$.

\subsubsection{$t_{\rm 2g}-t_{\rm 2g}$ contributions}

Introducing the effective hopping integral $t_{\mathrm{eff}} =  \dfrac{t_0^2}{3}\left( \dfrac{2}{\Delta-\lambda_p/2} +\dfrac{1}{\Delta+ \lambda_p} \right)$ between $M_1$ and $M_2$ via $p$-orbitals, and the ratio $r = \dfrac{2 \lambda_p}{2 \Delta +\lambda_p}$ between SOC $\lambda_p$ and the atomic energy difference $\Delta$, the hopping matrix involving only $t_{\rm 2g}$ orbitals, denoted by $t_0$ in Fig.~\ref{fig:conventions}, can be simplified in block form to
\begin{equation}\label{eq:t2gt2g_eff}
\mathbf{T^{eff}_{M_1M_2}}(t_{\rm 2g}\otimes t_{\rm 2g}) =
t_{\mathrm{eff}}\left(\arraycolsep=5pt\def\arraystretch{2}\begin{array}{rrr}
                    0_{2\times2} &                  \mathbf{\sigma}_o &  i \dfrac{r}{2} \mathbf{\sigma}_{x} \\
                 \mathbf{\sigma}_{o} &                   0_{2\times2} & -i \dfrac{r}{2} \mathbf{\sigma}_{y} \\
 -i \dfrac{r}{2} \mathbf{\sigma}_{x} & i \dfrac{r}{2} \mathbf{\sigma}_{y} &                    0_{2\times2}         
\end{array}\right),
\end{equation}
where $t_{\rm 2g}=\{d_{yz},d_{xz},d_{xy}\}$ and $\sigma_i$ with $i=x,y,z$ are the Pauli matrices carrying the spin degrees of freedom and $\sigma_o$ is  the $2\times 2$ identity matrix.
The holes in the intermediate sates at the X site and their indirect hopping $\mathbf{T_{M_i X_m}}$ determine the type of $\sigma$ matrices in the effective hopping 
For example in the limit $\lambda_p \rightarrow 0$, only the $d_{yz}-d_{xz}$ term $t_{\mathrm{eff}}\sigma_0 \rightarrow  t^2_0/\Delta$ is present which contributes to the direct hopping channel.
The new terms present for non-zero $\lambda_p$ are the spin-flip (SF) terms between $d_{yz}/d_{xz}$ and $d_{xy}$. 
Such terms would normally not appear in the second order perturbation process, as they involve a $d_{yz}/d_{xz}-p_z$ hopping followed by a $p_x/p_y-d_{xy}$ hopping, which will only occur if $p_z$ is entangled with $p_x/p_y$. 
The SOC among the $p_x,p_y,p_z$ generates such entanglement. Furthermore, when SOC is the dominant energy scale of the hole states, the
wavefunctions are inevitably mixtures of $p$-orbitals and their spin, leading to $\sigma_i$ dependence proportional to the $r$ ratio of the SOC and atomic energy difference. 

The superexchange process involving only $t_{\rm 2g}$ orbitals Eq.~(\ref{eq:t2gt2g_eff}) result in
\begin{equation}\label{eq:spin_model_t2gt2g}
\arraycolsep=1.4pt\def\arraystretch{3}
\begin{array}{c}
J_0^{\rm t_{2g}} = \dfrac{8 t^2_{\mathrm{eff}}}{9 \left(U+2 J_H\right)} , \ K_0^{\rm t_{ 2g}}  = -\dfrac{4 (r t_{\mathrm{eff}})^2}{9 \left(U+2 J_H\right)}. 
\end{array}
\end{equation}

The spin-dependent hoppings have generated a $S^z_iS^z_j$ Kitaev interaction.
This can be rudimentarily understood by the following steps.
 To simplify the steps, we focus on one spin 1/2 electron hopping along the Z bond,
 through only a SF hopping. Imagining two sites starting in $(\uparrow,\uparrow)$ state,
  the SF hopping can lower the energy by the process:
$(\uparrow ,\uparrow)-\text{SF}-(0,\downarrow\uparrow)-\text{SF}-(\uparrow,\uparrow)$,
at a energy cost of $-(r t_{\mathrm{eff}})^2/U$.
On the other hand,  if the two sites start in $(\uparrow,\downarrow)$ the SF process is forbidden from Pauli exclusion principle. 
Thus the exchange path starting from $(\uparrow,\uparrow)$ and ending in $(\uparrow,\uparrow)$ lowers the energy, 
which generates the FM $S^z_iS^z_j$ interaction.  
Carrying out the details for the S=3/2 leads to the expression of the spin exchange terms shown in  Eq.~(\ref{eq:spin_model_t2gt2g}).

Among the symmetry allowed terms, the symmetric off-diagonal term with operator $S^x_iS^y_j + S^y_iS^x_j=\dfrac{i}{2}\left(S_i^-S_j^--S_i^+S_j^+\right)$ does not occur in the second order perturbation results. This operator would connect $(\uparrow,\uparrow)$ to $(\downarrow,\downarrow)$, which is definitely possible from a SF process, however, there is a subtle cancellation. Having a single Pauli matrix in the SF term creates such cancellation among the two paths 
within the spin block, resulting in a null $\Gamma$ term. Thus it becomes finite only when higher order perturbation terms are included, or
when the octahedra are no longer ideal, as we will show in the later Sec.~\ref{sec:distortions}.

\subsubsection{$e_{\rm g}-t_{\rm 2g}$ contributions}

Given that the $p$-orbital's hybridization with $e_{\rm g}$ is larger than with
$t_{\rm 2g}$, denoted by $t_1$ and $t_2$ in Fig.~\ref{fig:conventions},this contribution is essential.
The final form of the effective  $\mathbf{T^{\mathrm{eff}}_{M_1M_2}}(e_{\rm g}\otimes t_{\rm 2g})$ hopping includes SF terms in the $d_{yz},\ d_{xz}$ to $e_{\rm g}$ blocks, as well as spin-independent $-2t_{\rm eff}(t_2/t_0)\sigma_0$ hopping between $d_{xy}$ and $d_{3z^2-r^2}$. The effective hopping matrix involving $e_{\rm g}-t_{\rm 2g}$ hoppings, in block matrix form, reads

\begin{equation}\label{eq:egt2g_eff}
\mathbf{T^{eff}_{M_1M_2}}(e_{\rm g}\otimes t_{\rm 2g}) =
t_{\mathrm{eff}}\left(\arraycolsep=5pt\def\arraystretch{2}\begin{array}{rrr}
  i \dfrac{r}{2} \dfrac{t_1}{t_0}  \sigma_{y} &  i \dfrac{r}{2} \dfrac{t_1}{t_0}  \sigma_{x} &  \multicolumn{1}{c}{0_{2\times2}} \\
 -i  \dfrac{r}{2} \dfrac{t_2}{t_0}  \sigma_{y} &  i \dfrac{r}{2} \dfrac{t_2}{t_0}  \sigma_{x} & -2 \dfrac{t_2}{t_0} \sigma_{o} \\
\end{array}\right),
\end{equation}
where $e_{\rm g}=\{d_{x^2-y^2},d_{3z^2-r^2}\}$ and $t_{\rm 2g}=\{d_{yz},d_{xz},d_{xy}\}$. Carrying out the strong coupling expansion, these hoppings lead to the additional contribution to the Heisenberg and Kitaev interactions
\begin{equation}\label{eq:spin_model_egt2g}
\arraycolsep=1.4pt\def\arraystretch{3}
\begin{array}{l}
J_0^{\rm {e_g}} =-\dfrac{16 \; J_H\;  t_{\rm eff}^2}{3 \left(\Delta_c + U' - J_H\right)\left(\Delta_c + U' +3J_H\right)}\dfrac{t_2^2}{t^2_0},\\
K_0^{\rm {e_g} }=\dfrac{2 \; J_H (r \; t_{\rm eff})^2}{3 \left(\Delta_c + U' - J_H\right)\left(\Delta_c + U' +3J_H\right)}\dfrac{t_1^2+t_2^2}{t^2_0}.
\end{array}
\end{equation}
Similar to the $t_{\rm 2g}-t_{\rm 2g}$ case, the SF terms contribute to the Kitaev interaction, while 
the spin-independent hopping, $-2t_{\rm eff}(t_2/t_0)\sigma_0$, generates a FM Heisenberg term. 
The FM Heisenberg interaction originates from the competition of two exited states separated by Hund's coupling, i.e, $e_{\rm g}$ 
paths consistent with the earlier findings obtained by the first principle calculations.\cite{Kashin2020FMEg,Soriano2020CrX3}

\subsection{Direct hopping}
The direct hopping, denoted by $t_{d1}$, $t_{d2}$, $t_{d3}$ and $\tilde{t}_{d0}$ in $\mathbf{T_{M_1M_2}}$ between M$_1$ and M$_2$ is given by
\begin{equation}\label{eq:directhop}
\begin{array}{l}
\mathbf{T_{M_1M_2}}\left( t_{\rm 2g}\otimes t_{\rm 2g}\right)=\left(\begin{array}{ccccc}
 t_{d1} & t_{d2} & 0 \\
 t_{d2} & t_{d1} & 0 \\
 0      & 0      & t_{d3} \\
\end{array}\right)\otimes\mathbf{\sigma}_o,\\[0.7cm]
\mathbf{T_{M_1M_2}}\left(e_{\rm g}\otimes t_{\rm 2g}\right)=\left(\begin{array}{ccc}
0 & 0 & 0 \\
0 & 0 & \tilde{t}_{d0} \\
\end{array}\right)\otimes\mathbf{\sigma}_o.
\end{array}
\end{equation}
The Slater-Koster decomposition is explained in the Appendix.
Since there is no SF hopping terms, this does not generate the Kitaev interaction, but changes the Heisenberg interaction.

\subsection{Summary and Comments}

Combining both indirect and direct hopping contributions, the two exchange interactions for the ideal octahedra environment Eq.~(\ref{eq:spin_model_ideal}) are found to be
\begin{eqnarray}\label{eq:spin_model_all}
J_0 & = & \dfrac{4\left( 2t_{d1}^2+  2( t_{\rm eff} + t_{d2})^2 + t_{d3}^2 \right)}{9 \left(U+2 J_H\right)} \nonumber\\
&-&  \dfrac{4 J_H \left(2 t_{\rm eff} (t_2/t_0) - {\tilde t}_{d0} \right)^2}{3 \left(\Delta_c + U^\prime - J_H\right) \left(\Delta_c + U^\prime + 3 J_H\right)}, \nonumber\\
K_0 & = & -\dfrac{4\;  (r\; t_{\rm eff})^2}{9 \left(U+2 J_H\right)}  \nonumber\\
&+ &  \dfrac{2 \; J_H (r \; t_{\rm eff})^2}{3 \left(\Delta_c + U^\prime - J_H\right) \left(\Delta_c + U^\prime + 3 J_H\right)}\dfrac{t_1^2+t_2^2}{t^2_0}.
\end{eqnarray}
Thus for the ideal honeycomb structure, the $J-K$ model is derived within the second order perturbation theory.

Some comments are useful, which will also motivate further investigation on the effects of trigonal distortion presented in the next section.
Kitaev and Heisenberg interactions have been generated while the $\Gamma$ interaction has not appeared at second order due to a subtle cancellation. The Kitaev has been generated purely from SF hoppings and has a prefactor of $r^2 = (2\lambda_p/(2\Delta+\lambda_p))^2$, while the Heisenberg includes spin-independent hopping contributions. Nevertheless both $K$ and $J$ have two contributions which come with opposite signs: one from $t_{2g}$ only paths and the other from the paths involving $e_{\rm g}$. This leads to a smaller Kitaev compared to 
Heisenberg interaction, unless the contribution from $e_{\rm g}$ paths reduces the overall strength of the Heisenberg interaction.
So long as the crystal field spitting is not excessively large, the naturally larger $e_{\rm g}$ hoppings will drive the system to a FM Heisenberg interaction.

The FM $J-K$ model pins the moment along the cubic axis when the quantum fluctuations are taken into account\cite{AvellaPRB2015,GiniyatPRB2016}, not along the observed [111] direction in CrI$_3$.
We thus investigate if other interaction terms may be generated.
The $\Gamma$ interaction allowed by the symmetry can be finite if higher order perturbation terms are included.
However, one may ask if there are other interactions allowed by a slight distortion of the lattice within the second order perturbation theory without invoking higher order terms.
Indeed TMT materials do not have ideal octahedra, but have either rhombohedral $R\bar{3}$ or monoclinic $C2/m$ structures, and
their magnetism strongly depends on  structural differences and number of layers.\cite{Klein2018CrI3,Song2018Giant,Klein2019NP,Kim2019PNAS} 
Below we study the effects of distorted octahedra, which induce additional hopping integrals which were forbidden without the distortion.

\section{Effects of distortion: distorted octahdera}\label{sec:distortions}

\begin{figure}[t]
  \centering
\begin{overpic}[width=0.9\linewidth]{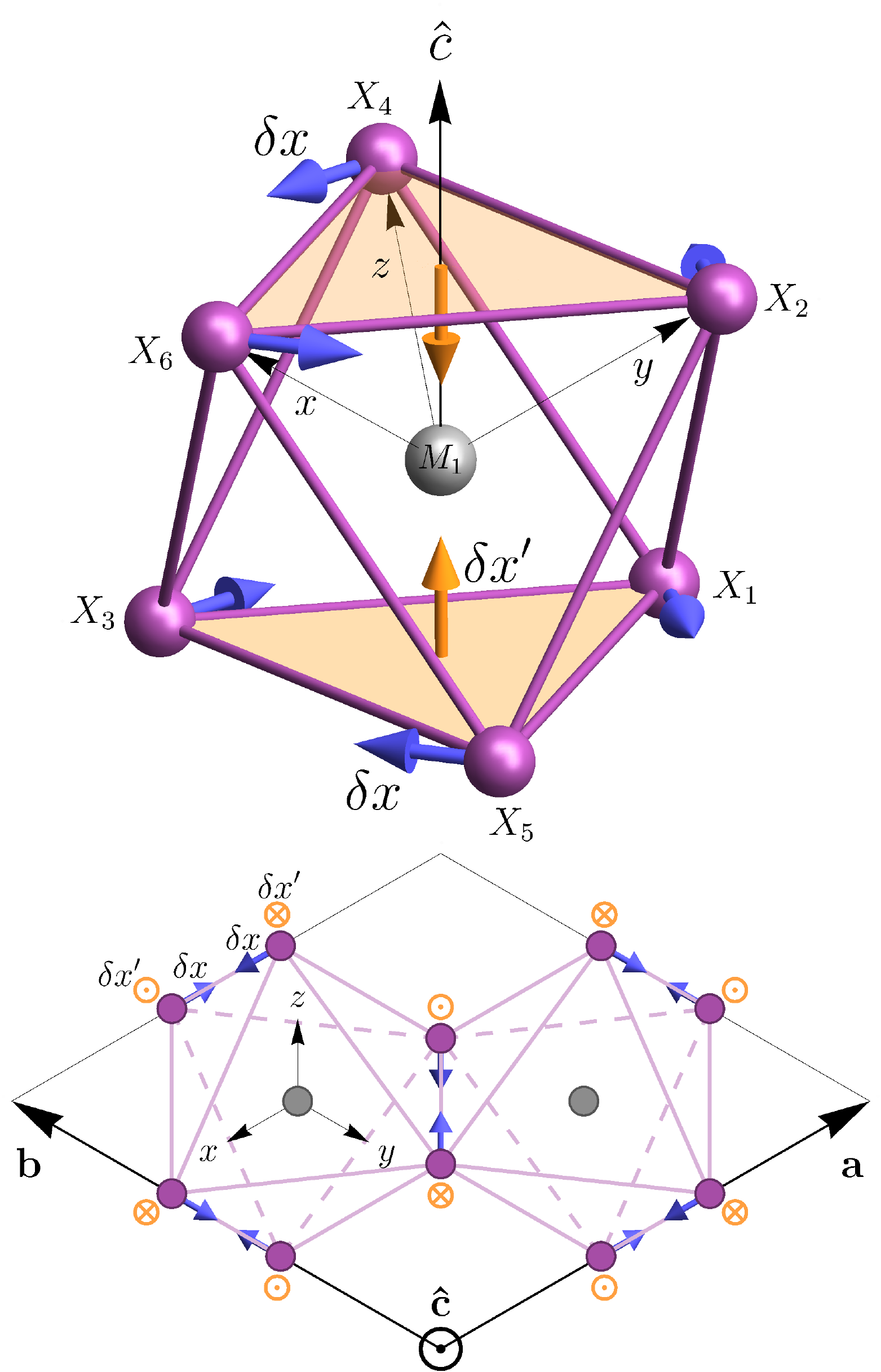}
\put(1,95){a)}
\put(1,40){b)}
\end{overpic}
  \caption{a) The distorted octahedra in $R{\bar 3}$ is shown. The octahedron made of X ligand can be viewed as two yellow triangles normal to the
  $\hat{c}=[1,1,1]/\sqrt{3}$ direction. The blue arrows represent new positions of X due to the staggered rotations of the two yellow triangular faces.
  The change of position X due to the staggered rotations is parameterized by $\delta x$.
  In addition, there is a compression of the two yellow triangles, squeezed as shown by the orange arrows parallel to the $\hat{c}$-axis. The change of position due to
  the compression is parameterized by $\delta x'$.  
  b) A top view of distortions in a unit cell is shown. The dotted circles $\odot$ indicate the new position of X moving out of the page along +$\hat{c}$ 
  while the circled cross $\otimes$ indicate a new position of X moving into the page toward $-\hat{c}$. 
  The exact positions of X$_{(1,2,3,4,5,6)}$ as a function of $\delta x$, $\delta x'$ are found in the Appendix.}
  \label{fig:octa_distortions}
\end{figure}

CrI$_3$ goes through a structural transition from $C2/m$ to $R{\bar 3}$ structure at low temperature.\cite{McGuire2015CoM}
In the rhombohedral structure, there are two types of X ligand position deviations from the ideal octahedra structure.
As shown in  Fig.~\ref{fig:octa_distortions}(a), a single octahedron can be viewed as two shaded triangles. 
One distortion is the staggered rotations of the two triangles denoted by $\delta x$ blue arrows, with displacements of X sites perpendicular to the $\hat{c}=[1,1,1]/\sqrt{3}$ direction. 
The other distortion is the compression of the distance between these two triangles along the $\hat{c}$-axis denoted by $\delta x'$ orange arrows, with displacements along the $\hat{c}$ direction. 
Here the dimensionless parameters $\delta x$ and $\delta x'$ are in units of the distance between the n.n. M sites $d_{M}$. Analytic formulas of the new positions of X sites under staggered rotations and compression are found in Appendix~\ref{apx:possitiondistortions}.
In the $R\bar{3}$ space group there are other types of distortions, namely the M$_1-$X$_{2,4,6}$ bond length can be different from the M$_1-$X$_{1,3,5}$ bond length. This type of distortion is generally exceedingly small compared to $\delta x$ and $\delta x'$ and we neglect it in the following analysis. 
Fig.~\ref{fig:octa_distortions}(b) shows a top view of the the honeycomb unit cell with two such distorted octahedra forming
 the Z bond, and the staggered rotation of the right octahedron is the mirror image of the left octahedron. 

The distortion-induced hopping matrices have all elements non-zero in $\mathbf{T_{M_iX_m}}$  as a result of lowering the local symmetry of the octahedron from $O_h$ to $D_3$.
We denote the new distortion allowed hopping integrals as $\delta t_i$. Starting with the distortion-induced hoppings, we follow the procedure described in Sec.~\ref{sec:ideal}, namely we use the distortion-induced $\mathbf{T_{M_iX_m}}$ matrices to derive the effective $\mathbf{T^{\mathrm{eff}}_{M_1M_2}}(t_{\rm 2g}\otimes t_{\rm 2g})$ and $\mathbf{T^{\mathrm{eff}}_{M_1M_2}}(e_{\rm g}\otimes t_{\rm 2g})$. Details of their form is deferred to Appendix~\ref{apx:distortedhop}. Treating the distortion-induced effective hoppings as a perturbation against the on-site interactions Eq.~(\ref{eq:kanamori}) the minimal n.n. spin model is finally given by
\begin{equation}\label{eq:spin_model_dist}
\begin{array}{l}
H  = \displaystyle \sum\limits_{\langle ij \rangle  \in \alpha \beta (\gamma)} \Big[J {\bf S}_i \cdot {\bf S}_j + K S_i^\gamma S_j^\gamma  + \Gamma(S_i^\alpha S_j^\beta +S_i^\beta S_j^\alpha)\\
\ \ \ \ \ \ \ \ \ \ \ \ \ \ \ \ \ +\ \Gamma^{\prime}(S_i^\alpha S_j^\gamma+S_i^\beta S_j^\gamma + S_i^\gamma S_j^\alpha +S_i^\gamma S_j^\beta) \Big] \\
\ \ \ \ \ \ \ \  + \displaystyle \sum_{i} A_c({\bf S}_i \cdot \mathbf{\hat{c}})^2,
\end{array}
\end{equation}
where $\alpha, \beta, (\gamma)$ refers to the $\gamma$ bond taking $\alpha$ and $\beta$ spin components \cite{rau2014prl,rau2014trigonal}, and $\mathbf{\hat{c}}=[1,1,1]/\sqrt{3}$.
In addition to the $J-K$ terms, 
two symmetric off-diagonal terms $\Gamma$ and $\Gamma^\prime$ have been generated, as well as the single-ion term $A_c$ allowed by the $C_3$ symmetry present
 on every site. $\Gamma$, $\Gamma^\prime$ and $A_c$ are proportional to distortion induced hoppings $\delta t$ as well as $r$, thus both distortion-induced hoppings and SOC is needed to bring rise to these terms. The form of the spin model $J-K-\Gamma-\Gamma^\prime-A_c$ to leading order in ${\delta t_i}$  as well as $e_{\rm g}$ contributions are listed in Table \ref{tab:allspinmodelterms} in the Appendix.
Note that both Heisenberg and Kitaev interactions are renormalized by the distortion, 
but the Heisenberg interaction has a linear term in $\delta t$, while Kitaev interaction does not.

\section{Application to CrI$_3$}\label{sec:dft}
To apply the above model to understand the magnetism in CrI$_3$, 
we use \textit{ab initio} calculations to obtain the microscopic parameters. The calculation is performed with Vienna \textit{ab initio} simulation package (VASP)\cite{Kresse_vasp_1993}. We use the Perdew-Burke-Ernzerhof (PBE) functional\cite{perdew_pbe_1996} in our calculations. 
The experimental bulk structure\cite{McGuire2015CoM} is used for the calculations. 
We use WANNIER90\cite{mostofi_wannier90_2008} to extract the hopping integrals.

From the density functional theory results, we estimate the crystal field splitting $\Delta_c=1253$~meV from on-site Cr $d$-orbitals as well as a crystal field splitting from the on-site I $p$-orbitals of $\Delta_p=528$~meV.
From Cr $d$- and I $p$- orbitals we extract the atomic energy difference between Cr and I $\Delta=2070$~meV. Finally we find the dominant indirect $p-d$ hopping integrals $t_0$, $t_1$ and $t_2$ and the direct $d-d$ hoppings parameters, listed in Table~\ref{tab:hopping_dom}, as well as the distortion induced $p-d$ hoppings, listed in Table~\ref{tab:hopping_distortions}. We verified that $t_0$, $t_1$ and $t_2$ obtained by the \textit{ab initio} calculation match well with the Slater-Koster expectations of Eq. (\ref{eq:sk_ind_dom}).

\begin{table}
  \begin{ruledtabular}
  \begin{tabular}{ccc|cccc}
   $t_0$ & $t_1$ & $t_2$ & $t_{d1}$ & $t_{d2}$ & $t_{d3}$ & $\tilde{t}_{d0}$ \\
   \hline
   590.1 & -992.13 & -558.3 & 44.67 & -41.26 & -147.44 & -20.75 \\
  \end{tabular}
  \end{ruledtabular}
  \caption{\label{tab:hopping_dom} Indirect and direct hoppings in units of meV obtained by \textit{ab initio} calculations. The hopping integrals are defined in Eq.~(\ref{eq:indirect}) and Eq.~(\ref{eq:directhop}).}
\end{table}

\begin{table}
  \begin{ruledtabular}
  \begin{tabular}{ccccccccc}
   $\delta t_1$ & $\delta t_2$ & $\delta t_3$ & $\delta t_4$ & $\delta t_5$ & $\delta t_1'$ & $\delta t_2'$ & $\delta t_3'$ & $\delta t_4'$ \\
   \hline
   2.34 & 21.76 & -22.81 & -61.74 & 65.86 & 70.97 & -43.50 & -49.41 & 29.65 \\
  \end{tabular}
  \end{ruledtabular}
  \caption{\label{tab:hopping_distortions} Distortion-induced indirect hoppings in units of meV obtained by \textit{ab initio} calculations. The hopping integrals are defined in Eq.~(\ref{eq:distTMX}).}
\end{table}

In the \textit{ab initio} calculations, we find a sizable crystal field spitting $\Delta_p$ on the ligand I sites, which we have not taken into account in the earlier analysis.
It is about $528$~ meV, which 
 is comparable to an estimated SOC parameter of $\lambda_p=630$~meV. 
 To capture its effects we revisit the effective hopping derivation, and add the crystal field spitting $\Delta_p$ on the X sites. 
 Then the energy level in Eq.~(\ref{eq:integrateout}) split into three levels as 
 $\Delta E_{a}=\Delta-\frac{\lambda}{2},\Delta+\frac{1}{4}\left(2\Delta_p+\lambda \pm\sqrt{\left(2\Delta_p-\lambda\right)^2+8\lambda^2}\right)$.
 While the holes in I sites are no longer pure total angular momentum states, the SOC is sizable enough to carry the spin 
entanglement through to the spin model. We found that second order perturbation theory, including the spitting $\Delta_p$ in the effective hoppings, does not change the form of the spin model Eq.~(\ref{eq:spin_model_dist}).
The analytic formulas would be vastly more complex in this case, so we proceed to a direct numerical evaluation of $J-K-\Gamma-\Gamma^\prime-A_c$ exchange terms using effective hoppings $\mathbf{T^{eff}_{M_1 M_2}}$ obtained by the {\it ab-inito} calculations listed in the Tables.

\begin{figure}[t]
  \centering
    \includegraphics[width=0.99\linewidth]{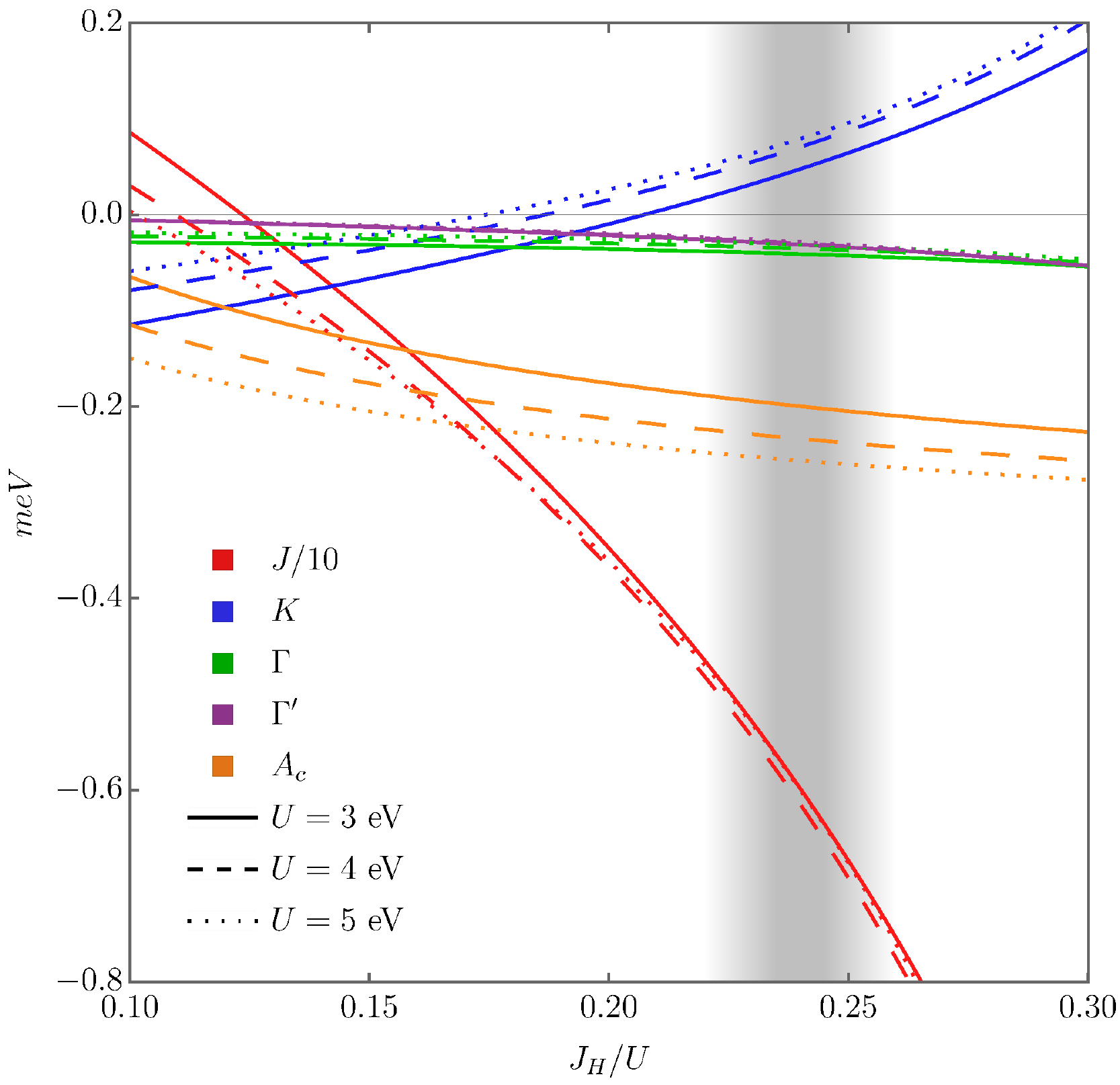}\\
  \caption{Spin model parameters $J$, $K$, $\Gamma$, $\Gamma^\prime$, and $A_c$ estimated from \textit{ab initio} parameters and plotted against $J_H/U$. Solid, dashed, and dotted lines correspond to $U=3{\rm eV},\ 4{\rm eV}\text{ and }5{\rm eV}$ respectively. Shaded region corresponds to $J_H/U\sim 0.24$ which is relevant to CrI$_3$ microscopics.}
  \label{fig:JKGGpAc_plot_from_dft}
\end{figure}

Assuming the spherical symmetry, i.e., $U^\prime=U-2J_H$,  we are left with two unknown parameters $U$ and $J_H$. 
We  plot $J$, $K$, $\Gamma$, $\Gamma^\prime$, and $A_c$ as a function of $J_H/U$ for several $U$ values, with results
shown in Fig.~\ref{fig:JKGGpAc_plot_from_dft}. 
The dominant term is Heisenberg except near the range $0.11\lesssim  J_H/U\lesssim 0.13$ where $J$ is almost zero before it changes the sign.
The sign of $J(K)$ are sensitive to the ratio $J_H/U$ and an adequately large Hund's coupling allows the FM Heisenberg (AFM Kitaev) interaction to persevere. 
This reflects the competition of the $t_{\rm 2g}$ vs $e_{\rm g}$ terms seen in Eq.~(\ref{eq:spin_model_all}), with $e_{\rm g}$ eventually wining over $t_{\rm 2g}$ leading to
the FM Heisenberg interaction. The distortion induced $\Gamma-\Gamma^\prime-A_c$ come in with FM sign, and the single-ion anisotropy $A_c$ is the sizable term.
Adopting the ratio of $J_H/U\sim0.24$ obtained by cRPA clculations~\cite{JangPRM2019}, CrI$_3$ sits in the shaded area in Fig.~\ref{fig:JKGGpAc_plot_from_dft}. CrI$_3$ is a ferromagnet with non-negligible $K-\Gamma-\Gamma^\prime-A_c$, and we now examine the implication of these exchange terms on the observed moment.


\begin{figure}[b]
  \centering
  \begin{overpic}[width=0.9\linewidth]{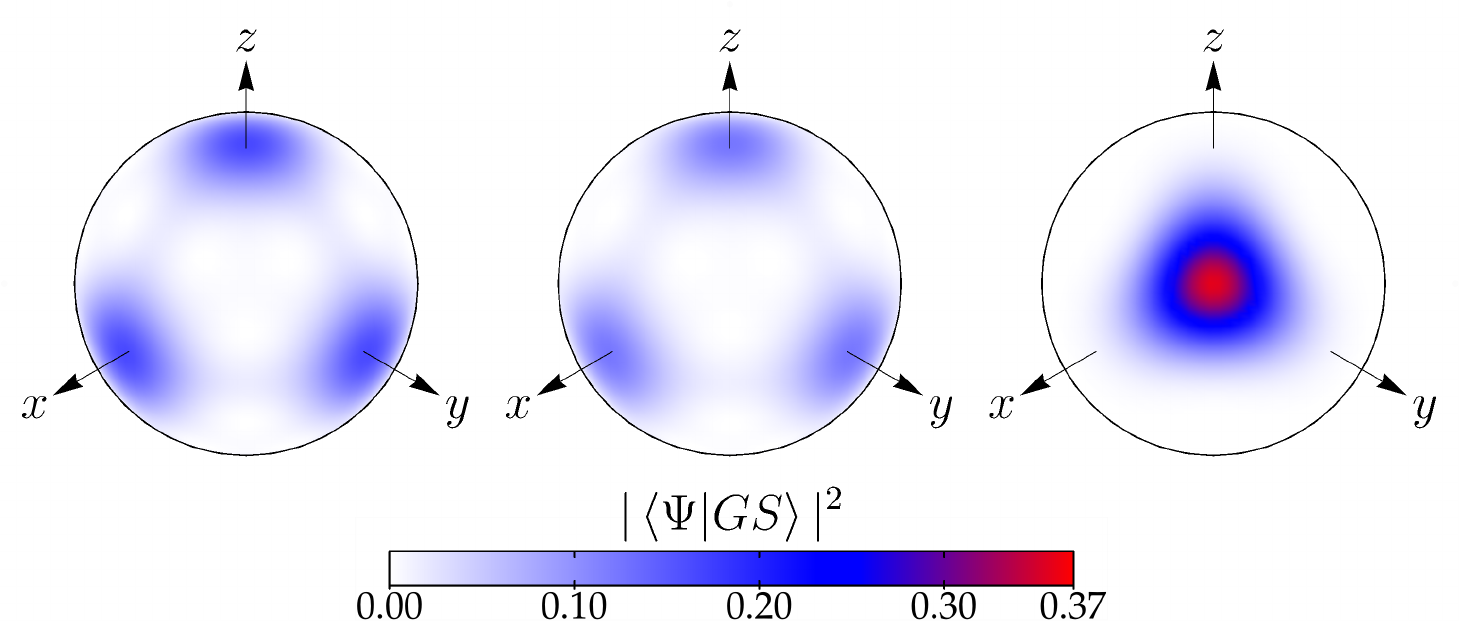}
\put(1,38){a)}
\put(34,38){b)}
\put(67.666,38){c)}
\end{overpic}
  \caption{Moment pinning calculation for difrent values of $(J,K,\Gamma)$ in the ferromagnetic phase: a) $(-1.00,-0.20,\ 0.00)$, b) $(-0.20,-1.00,\ 0.00)$, c) $(-0.20,-1.00,-0.02)$. Kitaev interaction always prefers the cubic axis in panel a) and b). A small interaction like $\Gamma$ is necessary to pin along the $\hat{c}$ direction as show in panel c).}
  \label{fig:moment1}
\end{figure}

\section{Magnetic Anisotropy}\label{sec:maganiso}
We have shown how the ideal honeycomb structure leads to $J-K$ model up to second order in perturbation, while distortions additionally generate $\Gamma-\Gamma^\prime-A_c$. 
The magnetic moment of the ferromagnetic state obtained with the FM $J-K$ model is pinned along the cubic axis 
such as $[100]$ and $C_3$ equivalent directions via quantum fluctuations\cite{AvellaPRB2015,GiniyatPRB2016}. We take a closer look at the effects of $\Gamma-\Gamma^\prime-A_c$ on the moment direction. We show three examples in Fig.~\ref{fig:moment1} to illustrate the moment pinning direction.

Following the method in Ref.~[\citenum{GiniyatPRB2016}], we perform exact diagonalization calculations on an eight site honeycomb cluster, with the cluster setup shown in the Appendix. Once the ground state wavefunction $\left| GS\right\rangle$ is obtained, 
the probability distribution $P= \left| \left\langle \Psi_{FM}(\theta,\phi) \right| \left. GS\right\rangle \right|^2$ is computed, where $\left| \Psi_{FM}(\theta,\phi) \right\rangle$ is a ferromagnetic ansatz with moment direction pointing at $(\theta,\phi)$ on the sphere. 
Results are show in Fig.~\ref{fig:moment1} for different values of $J$, $K$, and $\Gamma$. 
Independent of ratio of $J$ and $K$, the moment is along the cubic axis as shown in the panel (a) and (b).
On the other hand, when a small FM $\Gamma$ is introduced, the moment is along the $\hat{c}$ direction as shown in panel (c). 
This effect can be anticipated from the classical analysis. The classical $J-K-\Gamma$ model in the FM state with moment $\mathbf{S}_0 = (S^x_0,S^y_0,S^z_0)$ has an energy density per unit cell $u^{c}=(3J+K)+2\Gamma(S^x_0S^y_0+S^y_0S^z_0+S^z_0S^x_0)$, which demands $min\left[ u^{c} \right] \Rightarrow min \left[ sgn(\Gamma)(S^x_0S^y_0+S^y_0S^z_0+S^z_0S^x_0) \right]$. When $\Gamma<0$ we have $min \left[ -(S^x_0S^y_0+S^y_0S^z_0+S^z_0S^x_0) \right] \Rightarrow \mathbf{S}_0 = \hat{c}$ leading to a moment pined on the [1,1,1].  We choose $\Gamma/K \ll 1$ to show that a tiny FM $\Gamma$ anisotropy results in the ${\hat c}$-axis moment shown in Fig.~\ref{fig:moment1} (c).

FM $\Gamma^\prime$ has the identical effect as FM $\Gamma$ with respect to the moment direction of the ferromagnet. Clearly FM $A_c$ also promotes the ${\hat c}$-axis moment direction. All three interactions individually prefer the moment on the ${\hat c}$-axis. This effect is bigger than the quantum fluctuation effect of cubic axis pinning from the Kitaev term resulting in the observed ${\hat c}$-axis moment direction.

\section{Spin gap and finite transition temperature}\label{sec:spingap}

The preceding results suggest that it is likely that CrI$_3$ has dominant FM Heisenberg
and smaller Kitaev interaction, and 
the magnetic anistoropy originates from the distortion of the octahedra together with the SOC of heavy ligands.
The magnetic ordering pattern changes from bulk to films \cite{Huang2017MonolayerCrI3,SivadasNANO2018,JiangPRB2019,JangPRM2019,Kim2019PNAS,SorianoSSC2019}, suggesting a strong coupling between
magnetism and crystal structures, which further implies the importance of the distorted octahedra in the presence of SOC.
Our microscopic model has an interesting connection to the previous studies.
The Kitaev and single-ion anistoropy  in addition to the Heisenberg interaction were found in Ref.\cite{Xu2018Kitaev} using density functional theory. 
If $\Gamma = \Gamma^\prime$, the $J-\Gamma-\Gamma^\prime$ model
maps to the XXZ model\cite{Lado2017XXZ,Kim2019XXZ} in the $a-b-c$ crystallographic coordinate
\begin{eqnarray}
& J {\bf S}_i \cdot {\bf S}_j & + 
  \Gamma(S_i^x S_j^y +S_i^y S_j^x
 +  S_i^x S_j^z+S_i^z S_j^x + S_i^y S_j^z +S_i^z S_j^y) \nonumber\\
 &&  \Longrightarrow  \;\;\; J_{ab} {\bf S}_i \cdot {\bf S}_j + J_c  S^c_i S^c_j,
 \end{eqnarray}
where $J_{ab} = J -\Gamma$ and $J_c = 3 \Gamma$. 

Since the exchange parameters strongly depend on $J_H/U$, $\Delta_c/U$, and tight binding parameters, it is useful to obtain experimental inputs to determine some parameters. 
The inelastic neutron scattering experiments and magneto-Raman spectroscopy have reported a spin gap of approximately 0.36 meV at the Brillouin zone  (BZ) center
$\Gamma$-point.\cite{Chen2018Neutron,CenkerNP2020}
This is also consistent with a small anistoropy 
found in the ferromagnetic resonance experiment\cite{Lee2020PRL}, 
which is about $0.07$ meV leading to the spin gap of 0.3 meV.

Based on our spin wave analysis using the $J-K-\Gamma-\Gamma^\prime -A_c$ model including the second n.n. DM 
interaction, the spin wave dispersion $\omega_{\bf k}$ is expressed as  $\omega_0  + \rho \; k^2$ around the $\Gamma$-point.
Here $\omega_0$ and $\rho$ are the spin gap and stiffness, respectively, and they are given by 
\begin{equation}
\begin{gathered}
\omega_o  =  -S(3\Gamma+6\Gamma'+2A_c), \\
\rho  = \dfrac{S}{12} \left| 3J +K - \Gamma - 2\Gamma^\prime -\tfrac{(K+2\Gamma-2\Gamma^\prime)^2}{2\left(2\Gamma + 4\Gamma^\prime +2A_c+ 3J +K\right)} \right| 
\end{gathered}
\end{equation}
The details of the linear spin wave theory (LSWT) is presented in Appendix~\ref{apx:spinwave}.

Kitaev and DM interactions do not generate a gap at the $\Gamma$-point within the LSWT. The classical ferromagnetic ground state under the Kitaev and DM terms have a continuous degeneracy, and as a result expanding around this ground state in the LSWT will not result in a spin gap from these terms. The spin gap is rather small, as expected because it is originated from a combination of slightly distorted
octahedra and SOC, i.e., $\Gamma$, $\Gamma\rq{}$, and $A_c$. While small, it is essential for a finite $T_c$ in a single layer CrI$_3$.
In the FM ordered phase, at low temperatures the magnons are excited and their number is given by
$N_s (T)  = \int d^2 k \frac{1}{e^{\beta \omega_{\bf k} } -1}
= \frac{\pi}{\beta \rho} \int_{\beta \omega_0} \frac{dx}{e^x-1}.$
Without the spin gap $\omega_0$, $N_s$ diverges in two-dimension at any temperature except $T=0$, 
i.e., the celebrated Mermin-Wagner theorem\cite{MerminWagnerPRL1966}. 
Thus one can understand the essential role of $\omega_0$ which cuts the divergence,
and allows the FM ordering at finite temperatures, as long as $\omega_0 (T)$ remains finite for $T \le T_c$. 
While quantifying the transition temperature requires further analysis \cite{MaksimovPRR2020} ,
the temperature dependence of $\rho(T)$ and $\omega_0(T)$ from the inelastic neutron scattering measurement\cite{Chen2018Neutron} indicates the crucial role of $\omega_0(T)$ which vanishes at $T_c$.

Another important parameter is the Kitaev interaction $K$, which leads to a gap at the the BZ corner $K$-point known as the Dirac gap\cite{Lee2020PRL}, 
reported in the neutron scattering.\cite{Chen2018Neutron,Chen2020Neutron}
However, the second n.n. DM term
also generates the Dirac gap.\cite{Chen2018Neutron,Chen2020Neutron}
We would like to point out that $\Gamma$ and $\Gamma^\prime$ also play a part in the Dirac gap as shown in Eq.~(\ref{diracgap}).
Estimating the Kitaev interaction by an independent experimental measurement and further analysis on the individual role of the Kitaev and DM interactions remain to be resolved in future studies.

\section{Summary and Discussion} \label{sec:summary}

We have shown a microscopic derivation of the n.n. spin model for
honeycomb Mott insulators with three electrons in $t_{\rm 2g}$ orbitals at M sites surrounded by octahedral cages of ligands X with strong SOC.
Using the standard strong-coupling perturbation theory, we
found that there are only Heisenberg and Kitaev interactions for the ideal honeycomb lattice among
the three symmetry allowed interactions ($J, K, \Gamma$), because
$\Gamma$ is zero up to the second order perturbation.  The exchange paths between $t_{\rm 2g}$
and $t_{\rm 2g}$ vs. $t_{\rm 2g}$ and  $e_{\rm  g}$ via ligands
generate opposite signs for both Heisenberg and Kitaev interactions. The Heisenberg interaction is of order $t_{\rm eff}^2/U$, while
the Kitaev is smaller by a factor of $r^2 \sim(\lambda_p/\Delta)^2$.
The FM Heisenberg interaction originates from the $e_{\rm g}$ paths,
with the hopping integral between $e_{\rm g}$ and $p$-orbitals being larger compared to $t_{\rm 2g}$ and $p$-orbitals.

The FM Heisenberg and Kitaev interaction leads to FM ordering, but the moment direction is pinned
along the cubic $x$, $y$, or $z$ axis, e.g., [100] (and $C_3$ equivalent directions) via quantum fluctuations.
The $\hat{c}$-axis, [111], moment pinning found in CrI$_3$ should thus originate from other interactions,
which are also responsible for the spin gap at the $\Gamma$-point in the neutron scattering measurement\cite{Chen2018Neutron}.
Including the distorted octahedra present in the rhombohedral structure, three additional spin interactions are
found.  Inspecting the linear order in the distortion-induced hopping paths within the second order perturbation theory,
$\Gamma$, $\Gamma^\prime$ and single-ion anisotropy $A_c$ contain terms linear in
the distortion-induced hopping integrals.
The Heisenberg interaction also contains such additional linear term, but the Kitaev does not, implying that it is possible to 
fine-tune a system closer to the Kitaev dominant regime via trigonal distortions.

In this work we have focused on the nature of the monolayer, however, some comments are in place when considering the physics of the bulk. We showed that the $t_{\rm 2g}-t_{\rm 2g}$ vs. $e_{\rm g}-t_{\rm 2g}$ contributions to the intra-layer Heisenberg term $J$ are opposite in sign. This property should hold for inter-layer Heisenberg interaction $J_{\perp}$ as well, because it is determined from superexchange processes. Thus the magnetic ordering pattern between layers depends on the details of orbital compositions. From the \textit{ab initio} calculation, we found that the inter-layer hopping ranges from 10meV to 30meV. This leads to $J_{\perp}$ of order 0.1 meV. While it does not affect the intra-layer magnetism presented here, it is important for the ordering pattern in the bulk\cite{Huang2017MonolayerCrI3,SivadasNANO2018,JiangPRB2019,JangPRM2019,Kim2019PNAS,SorianoSSC2019}.


Comparison to $J_{\rm eff}=1/2$, S=1, and 3/2 spin systems would be useful.
The SOC is necessary to generate the bond-dependent interaction, as spin and orbitals should be entangled to get such a directional dependent spin interaction.
However, the presence of SOC is not enough to find an exotic phase like a spin liquid, because
the dominant interaction is often the Heisenberg interaction. 
To compare different spin cases, a summary of the ideal honeycomb exchange interactions, $J_0$, $K_0$, and $\Gamma_0$ including
the effective indirect hopping ($t_{\rm eff})$ and direct hopping ($t_d$) integrals, only up to the second order perturbation, is shown in the following Table \ref{tab:comparison}.

\begin{table}[ht!]
\arraycolsep=1.4pt\def\arraystretch{1.5}
\begin{tabular}{|c|c|c|c|}
\hline
&  & & \\[-0.4cm]
 Spin  & Heisenberg $J_0$ & \;\; Kitaev $K_0$ \;\; & \shortstack{symm. \\[-0.1cm] off-diagonal $\Gamma_0$}  \\
\hline
\hline
$J_{\rm eff} = 1/2 \; (d^5)$ & $O(\frac{t_d^2}{t_{\rm eff}^2})$   & $O(\frac{J_H}{U})$ & $O(\frac{t_d}{t_{\rm eff}}  \frac{J_H}{U})$ \\
\hline
$S=1 \;  (d^8)$   & $O(r^2)$ & $O(r^2)$ & 0  \\
\hline
$S=3/2 \; (d^3) $ & $O(1)$   & $O(r^2)$ & 0  \\
\hline
\end{tabular}
\caption{\label{tab:comparison} The leading term of the exchange interactions for the ideal honeycomb structure
in unit of $t_{\rm eff}^2/U$ for different spin $S$ including only up to the second order perturbation terms. 
See the Appendix for the full expression of $JK\Gamma\Gamma^\prime A_c$ for S=3/2 including the trigonal distortion-induced hopping contributions.}
\end{table}

Focusing on the ideal octahedra and n.n. model via second order superexchange processes,
$J_{\rm eff}=1/2$ is unique because the Heisenberg term is absent. On the other hand, for the S=1 model from $d^2$ in $e_{\rm g}$-orbitals,  the heavy ligand SOC generates
the Kitaev interaction, which  has the same order of magnitude with $J$. In fact, $K = - 2 J$, if only $e_{\rm g}$ paths are considered.\cite{Peter2019HigherK}
For S=3/2 case, we found that $J$ is order 1 in units of roughly $t_{\rm eff}^2/U$, while $K$ is smaller by $r^2$. Thus it is hard to compete with the Heisenberg
interaction. We speculate that this is valid for spins equal or higher than 3/2.  
Unlike S=1/2 and S=1 cases, the Heisenberg interaction is dominant in S=3/2, but a delicate cancellation among different contributions to the Heisenberg interaction,
may let the Kitaev interaction overtake a major place. In particular, the Heisenberg interaction is more sensitive to
the distortion-induced hopping integrals than the Kitaev term as shown in Table \ref{tab:allspinmodelterms} in the Appendix, 
manipulating ocathedra may be a way to tune the system to a desired Kitaev dominant regime.

In summary, in the ideal octahedra environment, we find that there are only two spin interactions, Heisenberg and Kitaev interactions.
Kitaev interaction is generally weaker compared to the Heisenberg interaction in contrast to the lower spin models. 
Indirect hoppings among $t_{\rm 2g}$ and $t_{\rm 2g}$ vs. $e_{\rm g}$ and $t_{\rm 2g}$ have opposite contributions. 
A detailed balance between the two indirect and direct hopping contributions highly depends on the hopping integrals, 
Hund\rq{}s coupling strength, and crystal field spitting.  $\Gamma$ interaction is absent up to the second order due to a subtle cancellation.
We further show that trigonal distortion allows three additional interactions, 
two symmetric off-diagonal interactions $\Gamma$ and $\Gamma^\prime$, and single-ion anisotropy $A_c$
along the $\hat{c}$-axis. They are all linearly proportional to a distortion-induced hopping integral. 
While they are much smaller than the Heisenberg interaction, 
they are essential for a spin gap in the FM phase of CrI$_3$ leading to a finite $T_c$.
Our study offers a microscopic route to the n.n. spin models, $J-K-\Gamma-\Gamma\rq{}-A_c$ interactions.
Given that there are five exchange terms within the n.n. model,
and second n.n. interactions including DM may be comparable to $\Gamma$, $\Gamma^\prime$ and $A_c$,
further theoretical and experimental studies are required to determine
the microscopic parameters of CrI$_3$ beyond the n.n. model.

->\begin{acknowledgments}
We acknowledge I. Lee, C. Hammel, N. Trivedi, P. Dai, and R. Valenti for useful discussion. H.Y.K. acknowledges
the funding from the Canada Research Chair Program. 
This work was supported by the Natural Sciences and Engineering Research Council of Canada, the Center for Quantum Materials at the University of Toronto, and the Canadian Institute for Advanced Research. Computations were performed on the Niagara supercomputer at the SciNet HPC Consortium. SciNet is funded by: the Canada Foundation for Innovation under the auspices of Compute Canada; the Government of Ontario; Ontario Research Fund - Research Excellence; and the University of Toronto.
\end{acknowledgments}

\bibliography{references_abriv}

\appendix*

\setcounter{equation}{0}
\renewcommand{\theequation}{A.\arabic{equation}}

\vspace{-5pt}
\section*{APPENDIX}\vspace{-10pt}

\subsection{Energy levels of on-site Kanamori Hamiltonian}\label{apx:spectrume}
\vspace{-5pt}

To obtain the n.n. spin interaction, we used the second order perturbation theory, where the dominant interaction is Eq.~(\ref{eq:kanamori}). 
First we note that in the lowest energy state there are three electrons at each metal site M. 
The exchange processes then involve one electron hopping between M sites. 
Thus the intermediate states have two electrons in $t_{\rm 2g}$-orbitals on one M site and four electrons in either $t_{\rm 2g}$- or $e_{\rm g}$-orbitals on the other M site. On the other hand the single-ion anisotropy would follow from a single M site, where the three electrons in the ground state interact with an exited three electron state. The energy levels of all states involved in these exchange processes are given in Table \ref{tab:spectrum}.
\begin{table}[t]
  \centering
\renewcommand{\arraystretch}{1.5}
\begin{tabular}{|ccl|}
 \hline
 Degen.        & Energy      &   \multicolumn{1}{c|}{Microscopics} \\
 \hline  
 \multicolumn{3}{|c|}{2 electrons (only in $t_{2g}$)} \\
  1 & $E_{A,2,1}$ & $=  U       +  2J_H + 2\epsilon_d$ \\
  2 & $E_{A,2,2}$ & $=  U       - \ J_H + 2\epsilon_d$ \\
  3 & $E_{A,2,3}$ & $=       U' + \ J_H + 2\epsilon_d$ \\
  9 & $E_{A,2,4}$ & $=       U' - \ J_H + 2\epsilon_d$ \\
 \hline 
 \multicolumn{3}{|c|}{3 electrons (GS)} \\
  4 & $E_{A,3,gs}$ & $= 3U' - 3J_H + 3\epsilon_d$ \\
 \multicolumn{3}{|c|}{3 electrons (only in $t_{2g}$)} \\
  4 & $E_{A,3,1}$ & $=       3U'         + 3\epsilon_d$ \\
  6 & $E_{A,3,2}$ & $=   U + 2U' -  2J_H + 3\epsilon_d$ \\
  6 & $E_{A,3,3}$ & $=   U + 2U'         + 3\epsilon_d$ \\
 \multicolumn{3}{|c|}{3 electrons (2-$t_{2g}$, 1-$e_{g}$)} \\
  8  & $E_{A,3,e1}$ & $=   U + 2U' -  2J_H + \Delta_c + 3\epsilon_d$ \\
  4  & $E_{A,3,e2}$ & $=   U + 2U' +   J_H + \Delta_c + 3\epsilon_d$ \\
  24 & $E_{A,3,e3}$ & $=       3U' -  3J_H + \Delta_c + 3\epsilon_d$ \\
  24 & $E_{A,3,e4}$ & $=       3U'         + \Delta_c + 3\epsilon_d$ \\
 \hline
 \multicolumn{3}{|c|}{4 electrons (only in $t_{2g}$)} \\
  1 & $E_{A,4,1}$ & $=  2U + 4U'         + 4\epsilon_d$ \\
  2 & $E_{A,4,2}$ & $=  2U + 4U' -  3J_H + 4\epsilon_d$ \\
  3 & $E_{A,4,3}$ & $= \ U + 5U' - \ J_H + 4\epsilon_d$ \\
  9 & $E_{A,4,4}$ & $= \ U + 5U' -  3J_H + 4\epsilon_d$ \\
 \multicolumn{3}{|c|}{4 electrons (3-$t_{2g}$, 1-$e_{g}$)} \\
  4  & $E_{A,4,e1}$ & $=       6U'         + \Delta_{c} + 4\epsilon_d$ \\
  6  & $E_{A,4,e2}$ & $=   U + 5U'         + \Delta_{c} + 4\epsilon_d$ \\
  10 & $E_{A,4,e3}$ & $=       6U' -  6J_H + \Delta_{c} + 4\epsilon_d$ \\
  18 & $E_{A,4,e4}$ & $=   U + 5U' -  4J_H + \Delta_{c} + 4\epsilon_d$ \\
  18 & $E_{A,4,e5}$ & $=       6U' -  2J_H + \Delta_{c} + 4\epsilon_d$ \\
  24 & $E_{A,4,e6}$ & $=   U + 5U' -  2J_H + \Delta_{c} + 4\epsilon_d$ \\
  \hline
\end{tabular}
\caption{\label{tab:spectrum} Spectrum of Hamiltonian Eq.~(\ref{eq:kanamori}). Listed states can contribute to the second order strong coupling perturbation.}
\vspace{-20pt}
\end{table}

\subsection{Ideal hoppings}\label{apx:idealhop}\vspace{-5pt}

\begin{figure}[t!]
  \centering
  \includegraphics[width=0.65\linewidth]{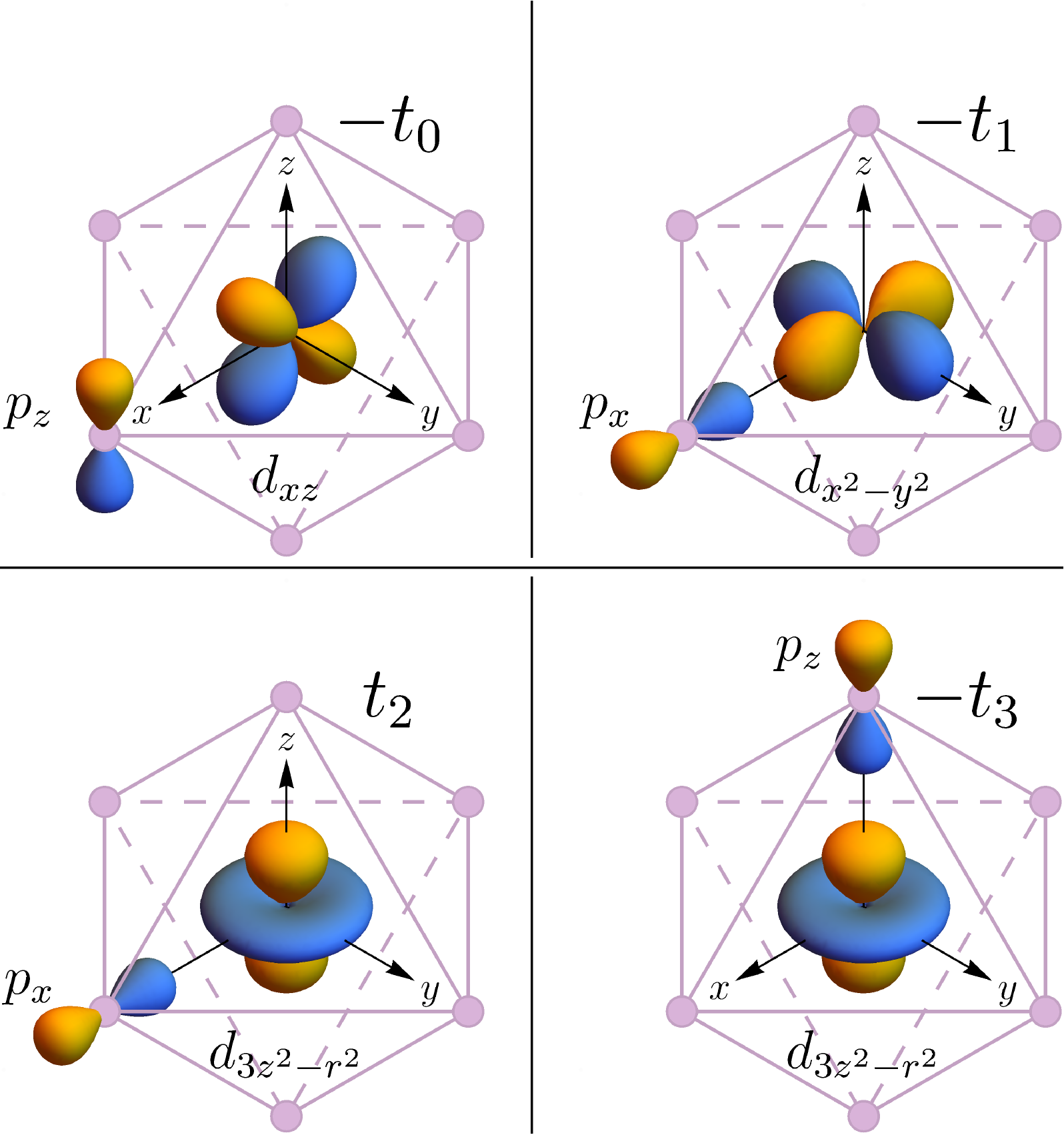}\\
  \caption{Indirect hoppings from X site to M site, in the global coordinates $xyz$.}
  \label{fig:hoppings}
\end{figure}

In the ideal octahedron environment the allowed indirect hoppings are show in Fig.~\ref{fig:hoppings}. On the M$_1-$X$_1$ bond we have the hopping matrix
\begin{equation}
\begin{array}{cccc}\label{eq:indirect}
& & \begin{array}{ccc} p_x & p_y & p_z \end{array} & \\[0.2cm]
\mathbf{T_{M_1X_1}} &=& \left(\begin{array}{ccc}
 t_1 &  0   &  0   \\
-t_2 &  0   &  0   \\
   0 &  0   &  0   \\
  0 &  0   &  t_0 \\
  0 &  t_0 &  0
\end{array}\right) & \begin{array}{l} d_{x^2-y^2} \\ d_{3z^2-r^2} \\ d_{yz} \\ d_{xz} \\ d_{xy} \ \ \ \ \ \ , \end{array} 
\end{array}
\end{equation}
The other indirect bonds $\mathbf{T_{M_2X_2}}, \ \mathbf{T_{M_2X_1}}, \ \mathbf{T_{M_1X_2}}$ are obtained from
 $\mathbf{T_{M_1X_2}}$ by consecutively applying the symmetry operation of the octahedron $C_4(0,0,1)$. Use of Eq.~(\ref{eq:indirect}) and symmetry related bonds are then used in Eq.~(\ref{eq:effective_total_hop}) to arrive at effective hoppings Eq.~(\ref{eq:t2gt2g_eff}) and Eq.~(\ref{eq:egt2g_eff}). Slater-Koster analysis\cite{SlaterKosterPR1954} allows us to decompose $t_0,\ t_1,\ t_2$ and $t_3$ into $\sigma$- and $\pi$-bonding integrals between $d$- and $p$-orbitals
\begin{align}\label{eq:sk_ind_dom}
\begin{array}{llll}
t_0 = t_{pd\pi},& t_1 = \dfrac{ \sqrt{3}t_{pd\sigma}}{2},& t_2 = \dfrac{ t_{pd\sigma}}{2}, & t_3 = t_{pd\sigma}, \\[0.3cm]
\multicolumn{4}{c}{t_{pd\sigma}<0,\ t_{pd\pi}>0,\ \left\lvert t_{pd\pi}\right\lvert < \left\lvert t_{pd\sigma}\right\lvert.}
\end{array}
\end{align}
The direct hoppings Eq.~(\ref{eq:directhop}) include $\delta$-bonding in addition to $\sigma$- and $\pi$-bondings
\begin{align}\label{eq:directdecompSK}
\arraycolsep=8pt\def\arraystretch{1.4}
\begin{array}{cc}
t_{d1}  =  \dfrac{ t_{dd\pi}    + t_{dd\delta}}{2} ,& t_{d2}  =  \dfrac{-t_{dd\pi}    + t_{dd\delta}}{2} \\
t_{d3}  =  \dfrac{3t_{dd\sigma} + t_{dd\delta}}{4} ,& \tilde{t}_{d0}  =  \dfrac{\sqrt{3}\left(t_{dd\sigma} - t_{dd\delta}\right)}{4}, \\[0.3cm]
\multicolumn{2}{c}{t_{dd\sigma}<0,\ t_{dd\pi}>0,t_{dd\delta}<0,} \\
\multicolumn{2}{c}{\left\lvert t_{dd\delta}\right\lvert < \left\lvert t_{dd\pi}\right\lvert < \left\lvert t_{dd\sigma}\right\lvert,}\\[0.3cm]
\end{array}
\end{align}
with the Anderse prediction $t_{dd\sigma}:t_{dd\pi}:t_{dd\delta}=-6:4:-1$.

\subsection{Positions of X sites under distortions}\label{apx:possitiondistortions}

In the $R{\bar 3}$ space group, octahedra made of two triangles as shown in Fig.~\ref{fig:octa_distortions} are rotated around ${\hat c}$-axis denoted by $\delta x$ blue arrows. Furthermore there is a slight change in the distance between these two triangles denoted by $\delta x'$ yellow arrows. 
The new positions of ligands are parameterized by dimensionless $\delta x$ and $\delta x'$, in units of the distance $d_M$ between M$_1$ and M$_2$. One can track the positions $\mathbf{r}_{m}$ of X$_m$, as it is related to the undistorted position $\mathbf{r}_{o,m}$, $\delta x$ and $\delta x'$:

\begin{align}\label{eq:Aposition}
\begin{array}{l}
\mathbf{r}_m=\mathbf{r}_{o,m}+\delta x\ d_{M} (\xi^a_m\ \hat{\mathbf{a}} + \xi^b_m\ \hat{\mathbf{b}} )+\delta x' \ d_{M} \zeta_m \ \hat{\mathbf{c}}\\[0.2cm]
\begin{array}{c|rrrrrrrrrrrrrr}
m & \multicolumn{5}{c}{\mathbf{r}_{o,m}} & & \multicolumn{5}{c}{\left( \xi^a_m,\xi^b_m \right)} & & \zeta_m \\
\hline
X_1 & ( & \frac{-d_M}{\sqrt{2}} & 0 & 0 & ) & & ( & -1 &, & -1 & ) & &  1 \\
X_2 & ( & 0 &  \frac{d_M}{\sqrt{2}} & 0 & ) & & ( &  1 &, &  1 & ) & & -1 \\
X_3 & ( & 0 & \frac{-d_M}{\sqrt{2}} & 0 & ) & & ( &  1 &, &  0 & ) & &  1 \\
X_4 & ( & 0 & 0 &  \frac{d_M}{\sqrt{2}} & ) & & ( & -1 &, &  0 & ) & & -1 \\
X_5 & ( & 0 & 0 & \frac{-d_M}{\sqrt{2}} & ) & & ( &  0 &, &  1 & ) & &  1 \\
X_6 & ( &  \frac{d_M}{\sqrt{2}} & 0 & 0 & ) & & ( &  0 &, & -1 & ) & & -1 \\
\end{array},
\end{array}
\end{align}
where X$_m$ lable sites as in Fig.~\ref{fig:octa_distortions}(a) and $\hat{\mathbf{a}}$, $\hat{\mathbf{b}}$ and $\hat{\mathbf{c}}$ are unit vectors along the lattice vectors of Fig.~\ref{fig:octa_distortions}(b).

\subsection{Distortion-induced hoppings}\label{apx:distortedhop}
To get the distortion-induced hoppings one needs to find the directional cosines of the indirect $d-p$ bonds using the ligand positions Eq.~(\ref{eq:Aposition}). With the metal site located at $\mathbf{r}_{\mathrm{M}_1}$ and ligand X$_m$ located at $\mathbf{r}_m$, the required directional cosines follow from $\mathbf{r}_{m}-\mathbf{r}_{\mathrm{M}_1}$. They are the inputs in the Slater-Koster formula for rotated bonds 
Ref.~[\citenum{SlaterKosterPR1954}].
Here we only show the leading term in each matrix element generated from the above procedure. Note that all elements become finite, and the distortion-induced hopping
between $t_{\rm 2g}$ and $p$-oribtials are denoted by $\delta t_i$ and $e_{\rm g}$ and $p$-orbitals by $\delta t^{\prime}_i$:

\begin{equation}\label{eq:distTMX}
\mathbf{T_{M_1X_1}}  = \left(
\begin{array}{ccc}
 t_1 & \delta t^{\prime} _1 & \delta t^{\prime} _2 \\
 -t_2  & \delta t^{\prime} _3 & \delta t^{\prime} _4 \\
 \delta t _1 & \delta t _2 & \delta t _3 \\
 \delta t _4 & \delta t _1 & t_0  \\
 \delta t _5 & t_0  & \delta t _1 \\
\end{array}
\right),
\end{equation}
where X$_1$ labels the site as in Fig.~\ref{fig:octa_distortions}(a). The distorted octahedron realizes the $D_3$ point group, which contains $C_3(1,1,1)$ and $C_2^\prime(-1,1,0)$ rotations. The hoppings $\mathbf{T_{M_1X_3}}$ and $\mathbf{T_{M_1X_5}}$ are recovered by applying $C_3(1,1,1)$ to $\mathbf{T_{M_1X_1}}$. Further $\mathbf{T_{M_1X_1}}$ relates to $\mathbf{T_{M_1X_2}}$ by a $C_2^{\prime}(-1,1,0)$. Finally $\mathbf{T_{M_1X_2}}$ relates to $\mathbf{T_{M_1X_4}}$ and $\mathbf{T_{M_1X_6}}$ by $C_3(1,1,1)$. The direct hopping integrals denoted by $\mathbf{T_{M_1 M_2}}$ is same the as Eq.~(\ref{eq:directhop}).
Making use of Eq.~(\ref{eq:distTMX}) we derive effective hoppings $\mathbf{T^{\mathrm{eff}}_{M_1M_2}}$ following the method in the main text Eq. (\ref{eq:effective_total_hop}). The distortion induced effective hoppings to leading order in $\delta t$ and listed in Table \ref{tab:effective_hop_dist}.

\begin{table*}[t]
\caption{\label{tab:effective_hop_dist} Effective hoppings to leading order in distortion-induced hoppings, where $\delta t$ are defined in Eq.~(\ref{eq:distTMX}). Distortion induced hoppings are grouped in: $   \delta t_a =  \delta t_1 - \delta t_4, \ 
    \delta t_b = 2\delta t_1 + \delta t_4, \ 
    \delta t_c =  \delta t_3 + \delta t_5, \ 
    \delta t_d =  \delta t_1 - \delta t_2, \
    \delta t_e =  \delta t_1 + \delta t_4, \
    \delta t_f =  \delta t_1 + \delta t_2 $.}
\begin{ruledtabular}
\begin{tabular}{ccc}
 & \\[-0.2cm]
$\mathbf{T^{\mathrm{eff}}_{M_1M_2}}(t_{\rm 2g} \otimes t_{\rm 2g})=t_{\rm eff} \left(\begin{array}{ccc}
T_{A}                & T_{C}                & T_D\\
T_{C}^{\dagger}      & T_{A}                & \ UT_DU^{\dagger}\\
T_D^{\dagger} & U^{\dagger}T_D^{\dagger}U & T_{B}
\end{array}\right)$ 
&
$\mathbf{T^{\mathrm{eff}}_{M_1M_2}}(e_{\rm g} \otimes t_{\rm 2g})=t_{\rm eff}\left(\begin{array}{ccl} 
\tilde{T}_{A}     & -U\tilde{T}_{A}U^{\dagger}  & \tilde{T}_{C}\\
\tilde{T}_{B}     & \ \ U\tilde{T}_{B}U^{\dagger}   & \tilde{T}_{D}\\
\end{array}\right)$ \\[0.6cm]
 \hline \\[-0.25cm]
 \multicolumn{2}{c}{$U = \mathrm{e}^{i  \pi \frac{1}{\sqrt{2}} \left( \frac{\sigma_{x}}{2} - \frac{\sigma_{y}}{2} \right) } = \dfrac{i}{\sqrt{2}}(\sigma_{x} - \sigma_{y})$}\\[0.4cm]
$\begin{array}{ccl}
    T_{A} &=&  \left(2\dfrac{\delta t_3}{t_0} +\dfrac{t_{d1}}{t_{\rm eff}} \right)  \sigma_{o} \\[0.3cm]
    T_{B} &=&  \left( 4 \dfrac{\delta t_5}{t_0} +\dfrac{t_{d3}}{t_{\rm eff}}\right)  \sigma_{o} \\[0.3cm]
    T_{C} &=& \left( 1 + \dfrac{t_{d2}}{t_{\rm eff}} \right)\sigma_{o}+i \dfrac{r}{2} \dfrac{\delta t_a}{t_0} \left( \sigma_{x}+ \sigma_{y}\right) \\[0.3cm]
    T_D   &=& \dfrac{\delta t_b}{t_0}   \sigma_{o} +i  \dfrac{r}{2}  \left( \sigma_{x}- \dfrac{ \delta t_c}{t_0} \sigma_{y} -\dfrac{\delta t_d}{t_0}  \sigma_{z} \right) \\[0.3cm]
\end{array}$
&
$\begin{array}{ccl} 
\tilde{T}_{A}   &=&   \left(\dfrac{\delta t^\prime_2}{t_0} + \dfrac{\delta t_d t_1}{t_0^2}  \right) \sigma_{o} 
   +i \dfrac{r}{2} \left( 
     \left(\dfrac{\delta t_3 t_1}{t_0^2} - \dfrac{\delta t^\prime_1}{t_0} \right) \sigma_{x}
   + \dfrac{t_1}{t_0} \sigma_{y}  
   - \dfrac{\delta t_e t_1}{t_0^2}  \sigma_{z} \right) \\[0.3cm]
\tilde{T}_{B}   &=&  \left(\dfrac{\delta t^\prime_4}{t_0} - \dfrac{\delta t_f t_2}{t_0^2}  \right) \sigma_{o} 
   +i \dfrac{r}{2} \left( 
     \left(\dfrac{\delta t_3 t_2}{t_0^2} - \dfrac{\delta t^\prime_3}{t_0} \right) \sigma_{x}
   - \dfrac{t_2}{t_0} \sigma_{y} 
   - \dfrac{\delta t_a t_2}{t_0^2}  \sigma_{z} \right)   \\[0.3cm]
\tilde{T}_{C}   &=&  i \dfrac{r}{2} \left(
   \left(\dfrac{\delta t_1 t_1}{t_0^2} 
   - \dfrac{\delta t_2^\prime}{t_0}\right)  \left( \sigma_{x} + \sigma_{y}\right) 
   -2 \left(\dfrac{\delta t_5 t_1}{t_0^2} - \dfrac{\delta t_1^\prime}{t_0}\right)  \sigma_{z} \right)\\[0.3cm]
\tilde{T}_{D} &=& \left(-2\dfrac{t_2}{t_0} +\dfrac{\tilde{t}_{d0}}{t_{\rm eff}}\right)  \sigma_{o} 
   + i \dfrac{r}{2} \left(\dfrac{\delta t_1 t_2}{t_0^2} + \dfrac{\delta t_4^\prime}{t_0}\right)  \left( \sigma_{x}- \sigma_{y}\right).   
 \end{array}$
    \end{tabular}
\end{ruledtabular}
\end{table*}

\begin{table*}[t]
\caption{\label{tab:allspinmodelterms} Spin model terms, under distortions, to leading order in $\delta t$. Distortion-induced $\delta t$ defined in Eq.~(\ref{eq:distTMX}), and character subscripted $\delta t$ defined in caption of Table \ref{tab:effective_hop_dist} and Eq.~(\ref{eq:SIAdist})}
\begin{ruledtabular}
\begin{tabular}{ccccc}
  &    &  &    &    \\[-0.2cm]
 &    & $t_{\rm 2g} \otimes t_{\rm 2g}$ &  & $e_{\rm g} \otimes t_{\rm 2g}$  \\[0.2cm]
 \hline
 & & & & \\
$J$ & $=$ & $\dfrac{4 \left(2 t_{d1}^2+ 2 ( t_{\rm eff} + t_{d2})^2 + t_{d3}^2 \right)}{9 \left(U+2 J_H\right)}$ 
& & $- \dfrac{4 J_H \left(2 t_{\rm eff} (t_2/t_0) - {\tilde t}_{d0} \right)^2}{3 \left(\Delta_c + U^\prime - J_H\right) \left(\Delta_c + U^\prime + 3 J_H\right)}$ \\
& & $+ \dfrac{32 t_{\rm eff}}{9 (U + 2 J_H)} \left( t_{d1} \dfrac{\delta t_3}{t_0} + t_{d3} \dfrac{\delta t_5}{t_0} \right)$ & \\[0.4cm]
$K$ & $=$ & $-\dfrac{4\;  (r\; t_{\rm eff})^2}{9 \left(U+2 J_H\right)} $
& & $\dfrac{2 \; J_H (r \; t_{\rm eff})^2}{3 \left(\Delta_c + U^\prime - J_H\right) \left(\Delta_c + U^\prime + 3 J_H\right)}\dfrac{t_1^2+t_2^2}{t^2_0} $\\[0.4cm]
$\Gamma$ & = & $-\dfrac{8 ( r t_{\rm eff})^2 }{9 \left(U+2 J_H\right)} \left( \dfrac{\delta t_c}{t_0} \right) $
& &  $- \dfrac{4 J_H  ( r t_{\rm eff})^2}{3 \left(\Delta_c + U^\prime - J_H\right) \left(\Delta_c + U^\prime + 3 J_H\right)} \left( \dfrac{ t_1 ( t_1  \delta t_3 - t_0 \delta t^\prime_1) }{t_0^3} - \dfrac{ t_2 ( t_2  \delta t_3 - t_0 \delta t^\prime_3)}{t_0^3} \right)$ \\[0.4cm]
$\Gamma'$ & =  & $-  \dfrac{4 (r t_{\rm eff})^2}{9 \left(U+2 J_H\right)} \left( \dfrac{\delta t_d}{t_0} \right)$
& &  $- \dfrac{2 J_H  ( r t_{\rm eff})^2}{3 \left(\Delta_c + U^\prime - J_H\right) \left(\Delta_c + U^\prime + 3 J_H\right)} 
\left( - \dfrac{ t_1^2 \delta t_e }{t_0^3} + \dfrac{ t_2^2 \delta t_a}{t_0^3} \right)$ \\[0.4cm]
$A_{c}$ & = &   $- \dfrac{4  (r t_{\rm eff})^2}{ \left(U+3 J_H-U'\right)}\left( \dfrac{\delta t_{\alpha}}{t_0} \right)$
& &  $-    \dfrac{4 J_H (r t_{\rm eff})^2}{ \Delta_{c} \left(\Delta_{c}+3 J_H\right)}  \left(  \dfrac{t_1}{t_0} \left( \dfrac{ \delta t_{\beta} t_1}{t_0^2} +\dfrac{\delta t^\prime_{\alpha}}{t_0}\right) -\dfrac{t_2 + t_3}{t_0} \left( \dfrac{ \delta t_{\beta} t_2  +  2 \delta t_1 t_3 }{t_0^2} +\dfrac{\delta t^\prime_{\beta}}{t_0} \right)  \right) $\\[0.4cm]
\end{tabular}
\end{ruledtabular}
\end{table*}

\subsection{Single-ion anisotropy term}
Due to the distortion and SOC, the single-ion anisotropic term is also generated via the hopping to anions which can hop back to create spin-dependent
on-site terms denoted by $\mathbf{T^{\rm eff}_{M_1 M_1}} = \mathbf{T^{\rm eff}_{M_2 M_2}}$. Without the trigonal distortion, the effective hopping integrals between $t_{\rm 2g}$ at M$_1$ are given by
\begin{align}
\arraycolsep=5pt\def\arraystretch{1.3}
\begin{array}{l}
\mathbf{T_{M_1M_1}^{\mathrm{eff}}}(t_{2g} \otimes t_{2g})= \\
t_{\mathrm{eff}} \left(
\begin{array}{rrr}
 4   \sigma_{o} & i r  \sigma_{z} & -i r  \sigma_{y} \\
 -i r  \sigma_{z} & 4   \sigma_{o} & i r  \sigma_{x} \\
 i r  \sigma_{y} & -i r  \sigma_{x} & 4   \sigma_{o} \\
\end{array}
\right).
\end{array}
\end{align}
Similarly, effective hopping between $t_{2g}$ and $e_g$ is found as
\begin{align}
\arraycolsep=5pt\def\arraystretch{2}
\begin{array}{l}
\mathbf{T_{M_1M_1}^{\mathrm{eff}}}(e_{g} \otimes t_{2g})= \\
t_{\mathrm{eff}} \left(
\begin{array}{rrr}
 i r \dfrac{t_1}{t_0}  \sigma_{x} & i r \dfrac{t_1}{t_0}  \sigma_{y} & -i 2r \dfrac{t_1}{t_0}  \sigma_{z} \\
 i r \dfrac{t_2+t_3}{t_0}  \sigma_{x} & -i r \dfrac{t_2+t_3}{t_0}  \sigma_{y} & \multicolumn{1}{c}{0_{2\times2}} \\
\end{array}
\right).
\end{array}
\end{align}
They lead to an ${\bf S}_i \cdot {\bf S}_i$ term, which is just equal to $S(S+1)$, an irrelevant constant to the spin model. Thus there is {\it no} single-ion anisotropy without the trigonal distortion.
With the trigonal distortion, the single-ion anisotropy along $\hat{c}$-axis is generated:
\begin{equation}\label{eq:SIAdist}
\begin{array}{ccl}
A_{c}  & = &  - \dfrac{4  (r t_{\rm eff})^2}{ \left(U+3 J_H-U'\right)}\left( \dfrac{\delta t_{\alpha}}{t_0} \right) \\[0.4cm]
 & & -    \dfrac{4 J_H (r t_{\rm eff})^2}{ \Delta_{c} \left(\Delta_{c}+3 J_H\right)}  \left(  \dfrac{t_1}{t_0} \left( \dfrac{ \delta t_{\beta} t_1}{t_0^2} +\dfrac{\delta t^\prime_{\alpha}}{t_0}\right)  \right.   \\[0.4cm]
& & \left. -\dfrac{t_2 + t_3}{t_0} \left( \dfrac{ \delta t_{\beta} t_2  +  2 \delta t_1 t_3 }{t_0^2} +\dfrac{\delta t^\prime_{\beta}}{t_0} \right)   \right),
\end{array}
\end{equation}
where $ \delta t_{\alpha} = - 2 \delta t_1 + \delta t_2 + \delta t_3 + \delta t_4 + \delta t_5$,
$\delta t_{\beta} = \delta t_2 + \delta t_3$,  $\delta t^\prime_{\alpha} = - \delta  t_1^\prime +\dfrac{\sqrt{3}\delta t_4^\prime -\delta t_2^\prime}{2}$, and $\delta t^\prime_{\beta} = - \delta  t_3^\prime + \dfrac{\sqrt{3}\delta t_2^\prime +\delta t_4^\prime}{2}$.

\subsection{Spin interaction with trigonal distortion}\label{apx:DistortedSpinModel}
Putting them all together, we have Heisenberg, Kitaev, $\Gamma$, $\Gamma^\prime$, and single-ion anisotropy $A_c$ term.
Their form to leading order in distortion induced hoppings $\delta t$ are summarized in Table \ref{tab:allspinmodelterms} bellow. Note that the Kitaev interaction does not have any linear term of distortion, and $\Gamma$, $\Gamma^\prime$
and single-ion anisotropy $A_c$ along $\hat{c}$-axis are finite due to both SOC and distortion.

\subsection{Spin Wave Theory}\label{apx:spinwave}

\begin{figure}[!h]
  \centering
\includegraphics[width=0.8\linewidth]{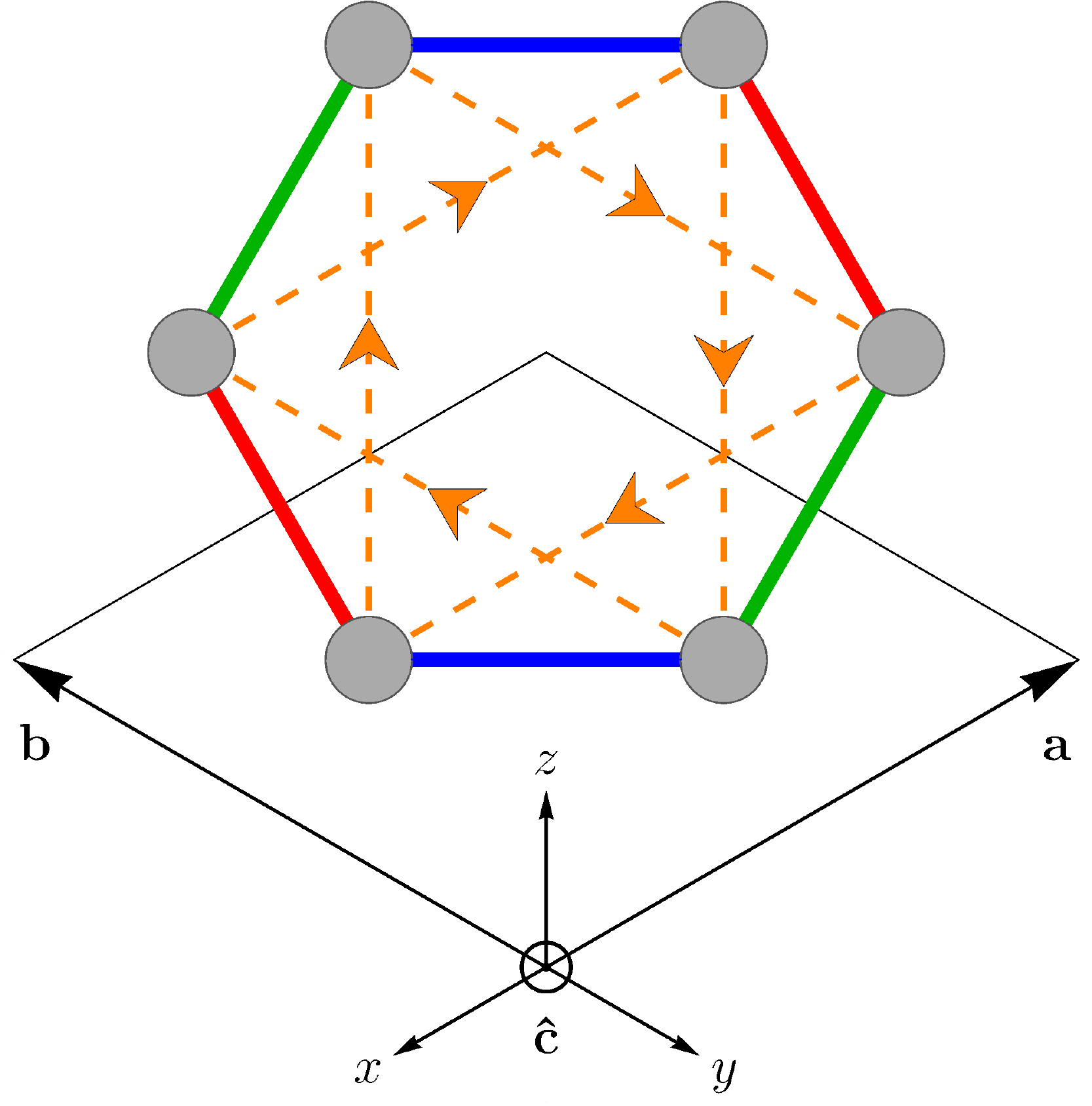}
  \caption{X Y and Z bonds of the $J-K-\Gamma-\Gamma^\prime-A_c$ shown in red, green and blue respectively. Second n.n are shown in orange. The orange arrows indicate when $\mathrm{sgn}(ij)=+1$ in the DM term.}
  \label{fig:swtmodel}
\end{figure}

We consider the $J-K-\Gamma-\Gamma^\prime-A_c$ as well as the second n.n. DM term ($\mathbf{D_c}$)

\begin{equation}\label{eq:SWTmodel}
\begin{array}{lcl}
H & = & \displaystyle \sum\limits_{ \substack{\langle i,j \rangle\in \alpha\beta(\gamma)} } \Big[ J\mathbf{S}_i\cdot\mathbf{S}_j + K S_i^{\gamma}S_j^{\gamma} + \Gamma\left( S_i^{\alpha}S_j^{\beta} + S_i^{\beta}S_j^{\alpha}\right) \\
& & \ \ \ \ \ \ \ \ \ \ \ \ \ \ \ + \Gamma'\left(S_i^{\alpha}S_j^{\gamma} + S_i^{\beta}S_j^{\gamma} + S_i^{\gamma}S_j^{\alpha} + S_i^{\gamma}S_j^{\beta} \right) \Big]\\[0.3cm]
& & +\displaystyle\sum\limits_{\langle\langle i,j \rangle\rangle} \mathbf{D_c} \cdot \left( \mathbf{S}_i\times\mathbf{S}_j \right) 
+\displaystyle\sum\limits_{i}  A_c \left(\mathbf{S}_i\cdot\mathbf{\hat{c}}\right)^2 ,
\end{array}
\end{equation}
where $\mathbf{D_c} = D_c\ \mathrm{sgn}(ij)\hat{c}$ and $\mathrm{sgn}(ij)=+1$ when $i$ to $j$ points along the orange arrows in Fig.~\ref{fig:swtmodel}.
The standard Holstein-Primakoff transformation~\cite{HolsteinPrimakoff} expanded to linear order in $S$ read
\begin{equation}
\begin{array}{lllll}
S^{+} &=& \sqrt{2S}\left(1-\dfrac{a^{\dagger}a}{2S}\right)^{\frac{1}{2}}a & \simeq &\sqrt{2S}a, \\[0.3cm]
\ S^{-} &=& \sqrt{2S}a^{\dagger}\left(1-\dfrac{a^{\dagger}a}{2S}\right)^{\frac{1}{2}} &\simeq& \sqrt{2S}a^{\dagger},\\[0.3cm]
\ S^{z} &=& S-a^{\dagger}a.
\end{array}
\end{equation}
Using the above and Fourier transforming Eq.~(\ref{eq:SWTmodel}) leads to
\begin{equation}
\begin{array}{c}
H=E_{CL} +  \sum\limits_{\mathbf{k}\in BZ}\mathrm{x}_\mathbf{k}^{\dagger} \mathbf{h_k} \mathrm{x}_\mathbf{k}, \\[0.3cm] 
\mathrm{x}_\mathbf{k}^{\dagger}=(a^{\dagger}_\mathbf{k},\ b^{\dagger}_\mathbf{k},\ a_\mathbf{-k},\ b_\mathbf{-k}), \\[0.5cm]
\mathbf{h_k} =\left(\begin{array}{cccc}
h_{o-}(\mathbf{k})     & h_{1}(\mathbf{k})      & 0                      & h_{2}(\mathbf{k})      \\
h_{1}(\mathbf{k})^*    & h_{o+}(\mathbf{k})   & h_{2}(-\mathbf{k})      & 0                      \\
0                      & h_{2}(-\mathbf{k})^*      & h_{o+}(\mathbf{k})      & h_{1}(\mathbf{k})      \\
h_{2}(\mathbf{k})^*      & 0                      & h_{1}(\mathbf{k})^*     & h_{o-}(\mathbf{k})
\end{array}
\right) ,
\end{array}\label{eq:bdg}
\end{equation}

\noindent
where the two species of bosons $a_\mathbf{k}$ and $b_\mathbf{k}$ correspond to the two sublattices of the unit cell. The $h(\mathbf{k})$ terms are

\begin{equation}
\begin{array}{l}
h_{o\pm}(\mathbf{k}) = h_{o} \pm h_{DM}(\mathbf{k}),  \\[0.2cm]
h_{o} = -S \left(2A_c+2\Gamma+4\Gamma'+3J+K\right),   \\[0.2cm]
h_{DM}(\mathbf{k}) = 2 S D_{c} \left( \sin({\bf a \cdot k }) + \sin({\bf b \cdot k }) \right.   \\
\ \ \ \ \ \ \ \ \ \ \ \ \ \ \ \ \ \ \ \ \ \ \ \ \ \ \ \ \ \ \ \ \ \ \ \ \left. - \sin({\bf (a + b) \cdot k })\right),  \\[0.2cm]
h_{1}(\mathbf{k}) = -\dfrac{S \left(\Gamma+2\Gamma'-3J-K\right)}{3}( 1+\mathrm{e}^{-i\mathbf{a}\cdot\mathbf{k}} +\mathrm{e}^{i\mathbf{b}\cdot\mathbf{k}}), \\[0.2cm]
h_{2}(\mathbf{k}) = \dfrac{S \left(2\Gamma-2\Gamma'+K\right)}{6}\left((1-i\sqrt{3})\mathrm{e}^{-i\mathbf{a}\cdot\mathbf{k}} \right.   \\ 
\ \ \ \ \ \ \ \ \ \ \ \ \ \ \ \ \ \ \ \ \ \ \ \ \ \ \ \ \ \ \ \ \ \ \left. + (1+i\sqrt{3})\mathrm{e}^{i\mathbf{b}\cdot\mathbf{k}}-2\right),
\end{array}
\end{equation}
where $\mathbf{a}$ and $\mathbf{b}$ are the lattice vectors in Fig.~\ref{fig:swtmodel}.
Following standard methods of diagonalizing BdG Hamiltonians\cite{COLPA1978327} we find the lowest eigenvalue around the $\Gamma$ point in the BZ, and upon expanding to orders of $k$, we get the spin gap $\omega_o$ and spin stiffness $\rho$:

\begin{equation}
\begin{array}{l}
\omega_{\mathbf{k}} = \omega_o + \rho k^2 \\[0.4cm]
\omega_o = S\left|3\Gamma+6\Gamma'+2A_c\right| \\[0.4cm]
\multicolumn{1}{l}{\rho = \dfrac{S}{12} \Bigg| 3J +K - \Gamma - 2\Gamma^\prime }\\ 
\multicolumn{1}{r}{\ \ \ \ \ \ \ \ \ \ \ \ \ -\dfrac{(K+2\Gamma-2\Gamma^\prime)^2}{2\left(2\Gamma + 4\Gamma^\prime +2A_c+ 3J +K\right)} \Bigg|.}
\end{array}
\end{equation}
At the K point in the BZ the Dirac gap is $\omega^K_+ - \omega^K_-$ where
\begin{equation}
\begin{array}{ccl}
\omega^K_+ &=& S\Big\lbrace\left(6\Gamma'+2A_c+3J\right)\left(4\Gamma+2\Gamma'+2A_c+3J+2K\right)  \\
   & & +6\sqrt{3}D_c\left( 2\Gamma+4\Gamma'+2A_c+3J+K\right) + 27 D_c^2\Big\rbrace^{1/2},  \\[0.4cm]
\omega^K_- &=& S\left|2\Gamma+4\Gamma'+2A_c+3J+K-3\sqrt{3}D_c\right| .
\label{diracgap}
\end{array}
\end{equation}

\subsection{Exact diagonalization calculations}\label{apx:ed}
\begin{figure}[!ht]
  \centering
  \includegraphics[width=0.8\linewidth]{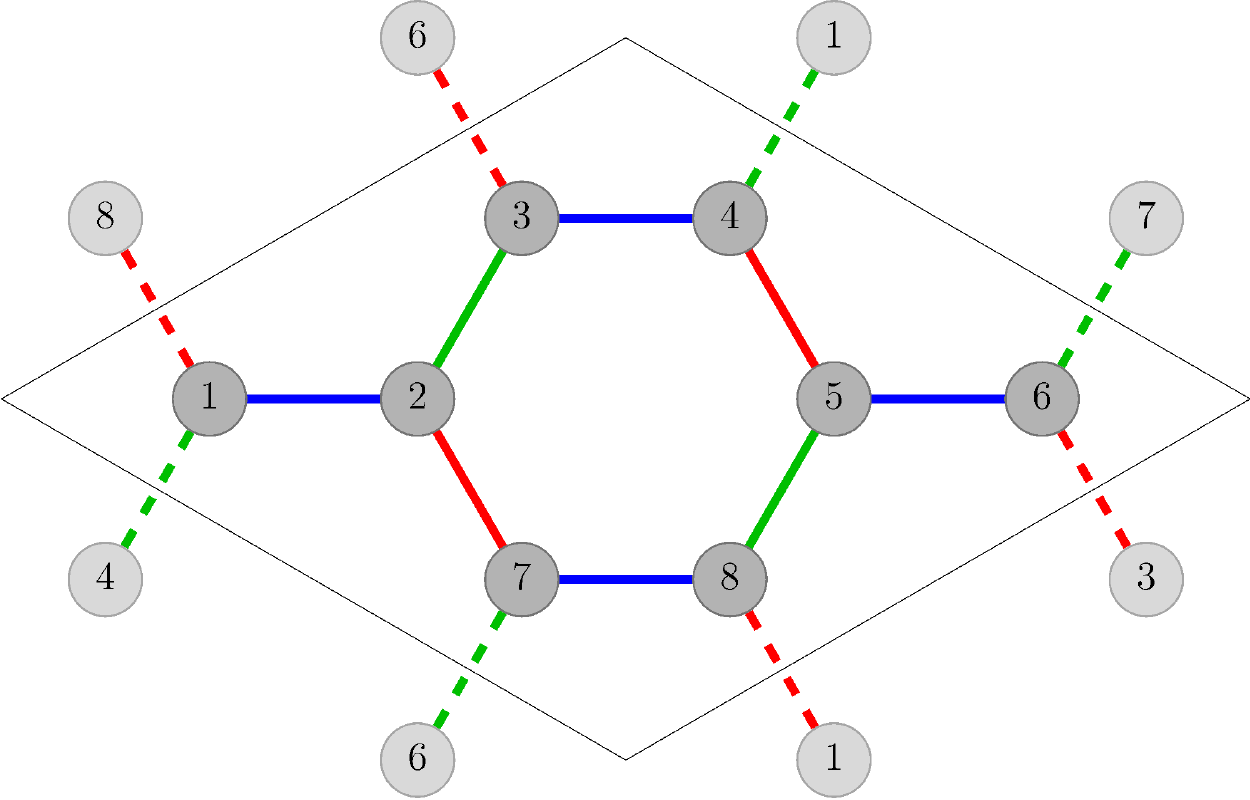}\\
  \caption{X,Y and Z bonds show in red, green and blue respectively. Dashed bonds represent the periodic boundary conditions.}
  \label{fig:edcluster}
\end{figure}

The moment pinning calculations are carried out by finding the ground state from exact diagonalization on an 8 site ($2\text{x}2$) honeycomb cluster with periodic conditions, shown in Fig.~\ref{fig:edcluster}.

\vspace{80pt}

\end{document}